\documentclass[twocolumn]{aastex631}
\usepackage{upgreek,xspace}

\newcommand{\Caltech}{\affil{California Institute of Technology, 1200 E. California Blvd., MC 249-17, Pasadena, CA 91125, USA}}

\newcommand{\GEMINI}{\affil{Gemini Observatory/NSF’s NOIRLab, 950 N. Cherry Avenue, Tucson, AZ, 85719, USA
}}

\newcommand{\OSU}{\affil{Department of Astronomy, The Ohio State University, 140 West 18th Avenue, Columbus, Ohio 43210, USA}}

\newcommand{\Alberta}{\affil{Department of Physics, University of Alberta, Edmonton, AB T6G 2E1, Canada}}

\newcommand{\ANU}{\affil{Research School of Astronomy and Astrophysics, Australian National University, Canberra, ACT 2611, Australia}}

\newcommand{\Carnegie}{\affil{Observatories of the Carnegie Institution for Science, 813 Santa Barbara Street, Pasadena, CA 91101, USA}}

\newcommand{\CfA}{\affil{Harvard-Smithsonian Center for Astrophysics, 60 Garden Street, Cambridge, MA 02138, USA}}

\newcommand{\CITEVA}{\affil{Centro de Astronomía (CITEVA), Universidad de Antofagasta, Avenida Angamos 601, Antofagasta, Chile}}

\newcommand{\ESO}{\affil{European Southern Observatory, Karl-Schwarzschild Stra{\ss}e 2, D-85748 Garching bei M\"{u}nchen, Germany}}

\newcommand{\Heidelberg}{\affil{Astronomisches Rechen-Institut, Zentrum f\"{u}r Astronomie der Universit\"{a}t Heidelberg, M\"{o}nchhofstra\ss e 12-14, D-69120 Heidelberg, Germany}}

\newcommand{\COOL}{\affil{Cosmic Origins Of Life (COOL) Research DAO, coolresearch.io}}

\newcommand{\ICRAR}{\affil{International Centre for Radio Astronomy Research, University of Western Australia, 35 Stirling Highway, Crawley, WA 6009, Australia}}

\newcommand{\ITA}{\affil{Universit\"{a}t Heidelberg, Zentrum f\"{u}r Astronomie, Institut f\"{u}r Theoretische Astrophysik, Albert-Ueberle-Str 2, D-69120 Heidelberg, Germany}}

\newcommand{\IWR}{\affil{Universit\"{a}t Heidelberg, Interdisziplin\"{a}res Zentrum f\"{u}r Wissenschaftliches Rechnen, Im Neuenheimer Feld 205, D-69120 Heidelberg, Germany}}

\newcommand{\JHU}{\affil{Department of Physics and Astronomy, The Johns Hopkins University, Baltimore, MD 21218, USA}}

\newcommand{\MPE}{\affil{Max-Planck-Institut f\"{u}r extraterrestrische Physik, Giessenbachstra{\ss}e 1, D-85748 Garching, Germany}}

\newcommand{\MPIA}{\affil{Max-Planck-Institut f\"{u}r Astronomie, K\"{o}nigstuhl 17, D-69117, Heidelberg, Germany}}

\newcommand{\OAN}{\affil{Observatorio Astron\'{o}mico Nacional (IGN), C/Alfonso XII, 3, E-28014 Madrid, Spain}}

\newcommand{\UToledo}{\affil{University of Toledo, 2801 W. Bancroft St., Mail Stop 111, Toledo, OH, 43606}}

\newcommand{\UBonn}{\affil{Argelander-Institut f\"ur Astronomie, Universit\"at Bonn, Auf dem H\"ugel 71, 53121 Bonn, Germany}}

\newcommand{\UChile}{\affil{Departamento de Astronom\'{i}a, Universidad de Chile, Camino del Observatorio 1515, Las Condes, Santiago, Chile}}

\newcommand{\UCSD}{\affil{Center for Astrophysics and Space Sciences, Department of Physics,  University of California,\\ San Diego, 9500 Gilman Drive, La Jolla, CA 92093, USA}}

\newcommand{\ULyon}{\affil{Univ Lyon, Univ Lyon 1, ENS de Lyon, CNRS, Centre de Recherche Astrophysique de Lyon UMR5574,\\ F-69230 Saint-Genis-Laval, France}}

\newcommand{\UWyoming}{\affil{Department of Physics and Astronomy, University of Wyoming, Laramie, WY 82071, USA}}

\newcommand{\UGent}{\affil{Sterrenkundig Observatorium, Universiteit Gent, Krijgslaan 281 S9, B-9000 Gent, Belgium}}

\newcommand{\STScI}{\affil{Space Telescope Science Institute, 3700 San Martin Drive, Baltimore, MD 21218, USA}}

\newcommand{\STScIESA}{\affiliation{AURA for the European Space Agency (ESA), Space Telescope Science Institute, 3700 San Martin Drive, Baltimore, MD 21218, USA}}

\newcommand{\INAF}{\affil{INAF -- Osservatorio Astrofisico di Arcetri, Largo E. Fermi 5, I-50157, Firenze, Italy}}

\newcommand{\SAIMSU}{\affil{Sternberg Astronomical Institute, Lomonosov Moscow State University, Universitetsky pr. 13, 119234 Moscow, Russia}}

\newcommand{\StockholmOKC}{\affil{The Oskar Klein Centre for Cosmoparticle Physics, Department of Physics, Stockholm University, AlbaNova, Stockholm, SE-106 91, Sweden}}

\newcommand\ionized[2]{[\mathrm{#1}\,\textsc{#2}]}
\newcommand\OII[1][]{\ifthenelse{\equal{#1}{}}{\ionized{O}{ii}}{\ionized{O}{ii}\,\uplambda#1}}
\newcommand\OIII[1][]{\ifthenelse{\equal{#1}{}}{\ionized{O}{iii}}{\ionized{O}{iii}\,\uplambda#1}}
\newcommand\NII[1][]{\ifthenelse{\equal{#1}{}}{\ionized{N}{ii}}{\ionized{N}{ii}\,\uplambda#1}}
\newcommand\SII[1][]{\ifthenelse{\equal{#1}{}}{\ionized{S}{ii}}{\ionized{S}{ii}\,\uplambda#1}}
\newcommand\SIII[1][]{\ifthenelse{\equal{#1}{}}{\ionized{S}{iii}}{\ionized{S}{iii}\,\uplambda#1}}
\newcommand\HA[1][]{\ifthenelse{\equal{#1}{}}{\mathrm{H}\,\upalpha}{\mathrm{H}\,\upalpha\,\uplambda#1}}
\newcommand\HB[1][]{\ifthenelse{\equal{#1}{}}{\mathrm{H}\,\upbeta}{\mathrm{H}\,\upbeta\,\uplambda#1}}

\newcommand\RA[4]{$#1^\mathrm{h}#2^\mathrm{m}#3^\mathrm{s}\kern -3pt.#4$}
\newcommand\DEC[4]{$#1^\mathrm{d}#2^\mathrm{m}#3^\mathrm{s}\kern -3pt.#4$}

\newcommand\HII{H\,\textsc{ii}\xspace}
\newcommand\HI{H\,\textsc{i}\xspace}

\let\tablenum\relax
\usepackage{siunitx}    

\DeclareSIUnit\arcsec{arcsec}
\DeclareSIUnit\arcmin{arcmin}
\DeclareSIUnit\parsec{pc}
\DeclareSIUnit\lightyear{ly}
\DeclareSIUnit\year{yr}
\DeclareSIUnit\mag{mag}
\DeclareSIUnit\erg{erg}
\DeclareSIUnit\Msun{M_\odot}
\DeclareSIUnit\Lsun{L_\odot}

\turnoffeditone
\begin{document}

\title[PHANGS-JWST Dust Maps]{PHANGS-JWST First Results:  The Dust Filament Network of NGC\,628 and its Relation to Star Formation Activity}
\suppressAffiliations
\correspondingauthor{David A. Thilker}
\email{dthilker@jhu.edu}

\author[0000-0002-8528-7340]{David A. Thilker}
\JHU

\author[0000-0002-2278-9407]{Janice C. Lee}
\GEMINI

\author[0000-0003-1943-723X]{Sinan Deger}
\Caltech
\StockholmOKC

\author[0000-0003-0410-4504]{Ashley~T.~Barnes}
\UBonn

\author[0000-0003-0166-9745]{Frank Bigiel}
\UBonn

\author[0000-0003-0946-6176]{M\'ed\'eric~Boquien}
\CITEVA

\author[0000-0001-5301-1326]{Yixian Cao}
\affiliation{Max-Planck-Institut f\"ur Extraterrestrische Physik (MPE), Giessenbachstr. 1, D-85748 Garching, Germany}

\author[0000-0002-5635-5180]{M\'elanie Chevance}
\Heidelberg
\COOL

\author[0000-0002-5782-9093]{Daniel A. Dale}
\UWyoming

\author[0000-0002-4755-118X]{Oleg~V.~Egorov}
\Heidelberg
\SAIMSU

\author[0000-0001-6708-1317]{Simon C.~O. Glover}
\ITA

\author[0000-0002-3247-5321]{Kathryn Grasha}
\ANU

\author[0000-0001-9656-7682]{Jonathan~D.~Henshaw}
\affiliation{Astrophysics Research Institute, Liverpool John Moores University, 146 Brownlow Hill, Liverpool L3 5RF, UK}
\affiliation{Max-Planck-Institut f\"ur Astronomie, K\"onigstuhl 17, D-69117 Heidelberg, Germany}

\author[0000-0002-0560-3172]{Ralf S.\ Klessen}
\ITA
\IWR

\author[0000-0001-9605-780X]{Eric Koch}
\CfA

\author[0000-0002-8804-0212]{J.~M.~Diederik Kruijssen}
\COOL

\author[0000-0002-2545-1700]{Adam K. Leroy}
\OSU

\author[0000-0002-4089-1704]{Ryan A. Lessing}
\JHU

\author[0000-0002-6118-4048]{Sharon E. Meidt}
\UGent

\author[0000-0001-5965-3530]{Francesca~Pinna}
\affiliation{Max-Planck-Institut f{\"{u}}r Astronomie, K{\"{o}}nigstuhl 17, D-69117 Heidelberg, Germany}

\author[0000-0002-0472-1011]{Miguel Querejeta}
\OAN

\author[0000-0002-5204-2259]{Erik~Rosolowsky}
\Alberta

\author[0000-0002-4378-8534]{Karin M. Sandstrom}
\UCSD

\author[0000-0002-3933-7677]{Eva Schinnerer}
\MPIA

\author[0000-0002-0820-1814]{Rowan J. Smith}
\affiliation{Jodrell Bank center for Astrophysics, Department of Physics and Astronomy, University of Manchester, Oxford Road, Manchester M13 9PL, UK}

\author[0000-0002-7365-5791]{Elizabeth~J.~Watkins}
\Heidelberg

\author[0000-0002-0012-2142]{Thomas G. Williams}
\MPIA

\author[0000-0002-5259-2314]{Gagandeep S. Anand}
\STScI

\author[0000-0002-2545-5752]{Francesco Belfiore}
\INAF

\author[0000-0003-4218-3944]{Guillermo A. Blanc}
\Carnegie
\UChile

\author[0000-0003-0085-4623]{Rupali Chandar}
\UToledo

\author[0000-0002-8549-4083]{Enrico Congiu}
\UChile

\author[0000-0002-6155-7166]{Eric Emsellem}
\ESO
\ULyon

\author[0000-0002-9768-0246]{Brent Groves}
\ICRAR
\ANU

\author[0000-0001-6551-3091]{Kathryn Kreckel}
\Heidelberg

\author[0000-0003-3917-6460]{Kirsten~L.~Larson}
\STScIESA

\author[0000-0001-9773-7479]{Daizhong Liu}
\MPE

\author[0000-0002-0873-5744]{Ismael Pessa}
\MPIA
\affiliation{Leibniz-Institut f\"{u}r Astrophysik Potsdam (AIP), An der Sternwarte 16, 14482 Potsdam, Germany}

\author[0000-0002-3784-7032]{Bradley~C.~Whitmore}
\STScI

\begin{abstract}

PHANGS-JWST mid-infrared (MIR) imaging of nearby spiral galaxies has revealed ubiquitous filaments of dust emission in intricate detail.  We present a pilot study to systematically map the dust filament network (DFN) at multiple scales between 25--400 pc in NGC~628.  MIRI images at 7.7, 10, 11.3 and 21$\micron$ of NGC~628 are used to generate maps of the filaments in emission, while PHANGS-HST B-band imaging yields maps of dust attenuation features.  We quantify the correspondence between filaments traced by MIR thermal continuum / polycyclic aromatic hydrocarbon (PAH) emission and filaments detected via extinction / scattering of visible light; the fraction of MIR flux contained in the DFN; and the fraction of HII regions, young star clusters and associations within the DFN.  We examine the dependence of these quantities with the physical scale at which the DFN is extracted.  With our highest resolution DFN maps (25 pc filament width), we find that filaments in emission and attenuation are co-spatial in 40\% of sight lines, often exhibiting detailed morphological agreement; that $\sim$30\% of the MIR flux is associated with the DFN; and that 75--80\% of HII regions and 60\% of star clusters younger than 5 Myr are contained within the DFN.  However, the DFN at this scale is anti-correlated with looser associations of stars \edit1{younger than 5 Myr} identified using PHANGS-HST near-UV imaging.  We discuss the impact of these findings for studies of star formation and the ISM, and the broad range of new investigations enabled with multi-scale maps of the DFN. 

\end{abstract}


\newcommand{\msun}{M$_{\odot}$}
\newcommand{\HST}{\textit{HST}\,}
\newcommand{\JWST}{\textit{JWST}\,}
\newcommand{\Gaia}{\textit{Gaia}\,}

\section{Introduction}
Two overwhelming impressions from inspecting \JWST images of nearby galaxies are the sheer number of resolved stars seen in the near-IR (NIR) and\edit1{ }
the stunning degree of structured, filamentary mid-IR (MIR) emission originating from small dust grains and polycyclic aromatic hydrocarbons (PAHs) in the interstellar medium (ISM). \JWST provides the spatial resolution necessary to cleanly decompose the observed MIR dust emission into filament features, discrete compact sources, and a diffuse component throughout the Local Volume ($d \lesssim$11 Mpc) and beyond, as could only previously be done in the Local Group \citep{Hinz2004, Barmby2006, Verley2007, Verley2009}. This is of astrophysical importance not only because dust plays a central role in enabling star formation, but also 
hides the youngest clusters and star-forming regions from view at short wavelengths. In \HST optical multi-color imaging of star-forming galaxies, dust lanes stand out as 
highly structured attenuation features \citep{LaVigne2006, Dong2016} occasionally punctuated by \HII regions and clusters that have pierced the veil of their dusty natal molecular cloud.

The observation of such abundant organized extragalactic structure in dust emission and attenuation is tantalising because studies of the cold gas and dust in the Milky Way have revealed 
filaments \citep{Jackson2010} of length $>$100~pc that have even been dubbed the `bones' of the Milky Way's cold ISM \citep{Goodman2014, Ragan2014, Zucker2015, Soler2020}. Indeed, much recent work in the Milky Way points towards a view of the cold, dusty star-forming medium that is 
filamentary and multi-scale \citep{Hacar2022, Pineda2022, Zucker2018}, very different from the classical `spherical molecular cloud'.  Filamentary structure even persists at sub-pc, cloud-substructure scales \citep[e.g.][]{Andre2010,Andre2014} though, in the Local Volume extragalactic context, we are limited to studying larger filaments \citep[akin to those of][]{Syed2022}. The shift to a filament-centered paradigm implies that criteria for stability and fragmentation change, becoming a mass per unit length threshold rather than a more traditional Jeans mass argument.  All this filamentary structure seeds star formation and determines the rate and efficiency of collapse, and defines the medium that the stellar feedback is subsequently driven into, thereby determining how feedback drives the baryon cycle within galaxies.

A revolution in our view of the dust structure in nearby galaxies is underway, having overcome the barrier of resolution with the combined capabilities of \JWST and \HST.  This enhanced extragalactic perspective is a critical advance \edit1{because we only have observations of the Milky Way from within, making them subject to distance uncertainties and line-of-sight projection ambiguities. Milky Way based studies of dust filament structure do sample substantially smaller physical scales than even JWST can reach in nearby galaxies, and the use of distance-tagged \Gaia measurements to infer 3D extinction maps \citep[e.g.][]{Sale2018,Green2019,Lallement2019,LeikeEnslin2019}, combined with a focus on dust emission out of the plane, together helps mitigate confusion. Indeed, recent studies comparing the 3D filament structures in extinction with 2D emission maps have shown remarkable correspondence between dust tracers \citep[e.g.][]{Hottier2021,Zucker2021,Bialy2021,Dharmawardena2022a,Dharmawardena2022b}. Nevertheless, we emphasize that the area of the Milky Way disk probed by these techniques is quite limited ($d\leq3$~kpc). In external galaxies,} we can now quantify entire \edit1{dust} filament networks \edit1{(DFNs)\footnote{All of the first four PHANGS-JWST galaxies observed show prominent structuring of the disk into a dust filament network.  For this reason, we coin the acronym DFN with confidence that it will be frequently used.  The concept of a filament network is not new \citep[e.g.][]{Hacar2022,Andre2014,Smith2014b,Smith2016}, but MIRI has revealed such complex multi-scale structuring is commonplace.}} on scales ranging from the size of individual GMCs up to morphological features dominating entire galaxies and reveal their intimate connection with respect to star formation, feedback, and dynamical mechanisms.  This new era of dusty ISM cartography will leverage representative galaxy samples to provide systematic answers to: how the prevalence and properties of these filamentary features may depend on galactic environment; whether they universally form the backbone of the cold ISM, through comparison to CO maps; and how the joint dust and molecular gas distribution is related to structures like spiral arms and bars, contrasted to high-resolution pc-scale UV+IR tracers of star formation activity from \HST and \JWST.  Critically, the new observations can be directly compared to state-of-the-art galaxy simulations \citep[e.g.][]{Smith2014, Smith2020, Duarte-Cabral2015, Duarte-Cabral2017, Tress2020, Tress2021, Jeffreson2020}.

JWST resolution and depth are sufficient to recover filamentary dust emission features, at GMC-scales, analogous to those of the Milky Way in galaxies out to \edit1{roughly} the distance of the Virgo cluster.  This letter focuses on NGC\,628 (also known as Messier~74, `The Phantom Galaxy') an archetypal face-on SA(s)c galaxy \cite{Buta2015}.  NGC\,628 is nearby  \citep[$d$\,=\,9.84$\pm$0.03\,Mpc;][]{Anand+2021b,Anand+2021a}, star-forming (SFR\,=\,1.8\,$\pm$\,0.45~\msun \rm yr$^{-1}$), massive ($M_{*}$\,=\,2.2\,$\pm$\,$0.56\times10^{10}$~\msun) and viewed at low inclination ($i$\,$\sim$\,9$^{\circ}$\,$\pm$\,12$^{\circ}$, \citealp{Lang+2020}). NGC\,628 is part of a broader sample of 19 ``main sequence'' star-forming galaxies for which systematic, uniform surveys with \HST \citep{PHANGS-Hst}, ALMA \citep{Leroy+2021a}\edit1{,} VLT-MUSE \citep{Emsellem+2022} and now \JWST \citep{LEE_PHANGSJWST} have been carried out by the PHANGS (Physics at High Angular resolution in Nearby GalaxieS) collaboration.\footnote{https://sites.google.com/view/phangs/home} 

\begin{figure*}
\includegraphics[width=1.0\linewidth]{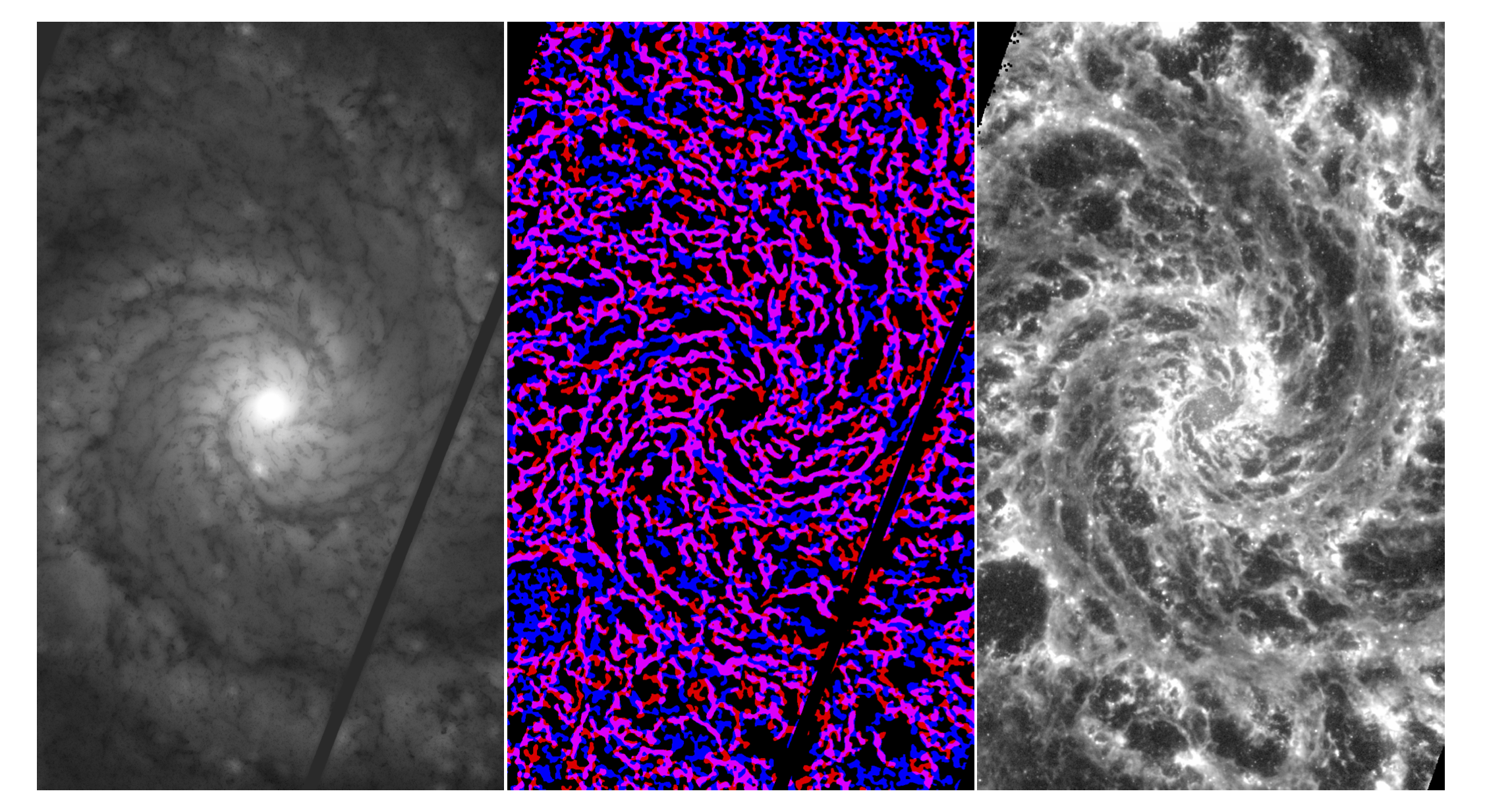}
\caption{\edit1{A first comparison of dust filaments identified in absorption and emission.} \textit{Left:} Pre-processed B-band \HST image, filtered to remove compact positive sources but retain small  negative/concave morphological features. \textit{Center:} Filament masks for visible attenuation (blue) and MIR emission (red) extracted at scale of 25~pc. Areas of overlap appear magenta. \textit{Right:} \JWST/MIRI F770W image. North is up, East is left, and the field of view spans 7.3 kpc from top to bottom. The figure only shows a portion of the area observed with \JWST. 
 The entire image may be seen in Figs.~\ref{fig:comparecumulativescales} and \ref{fig:appendixmasks1}--\ref{fig:appendixmasks4}.}
\label{fig:triofig}
\end{figure*}  

\begin{figure}
\includegraphics[width=\linewidth]{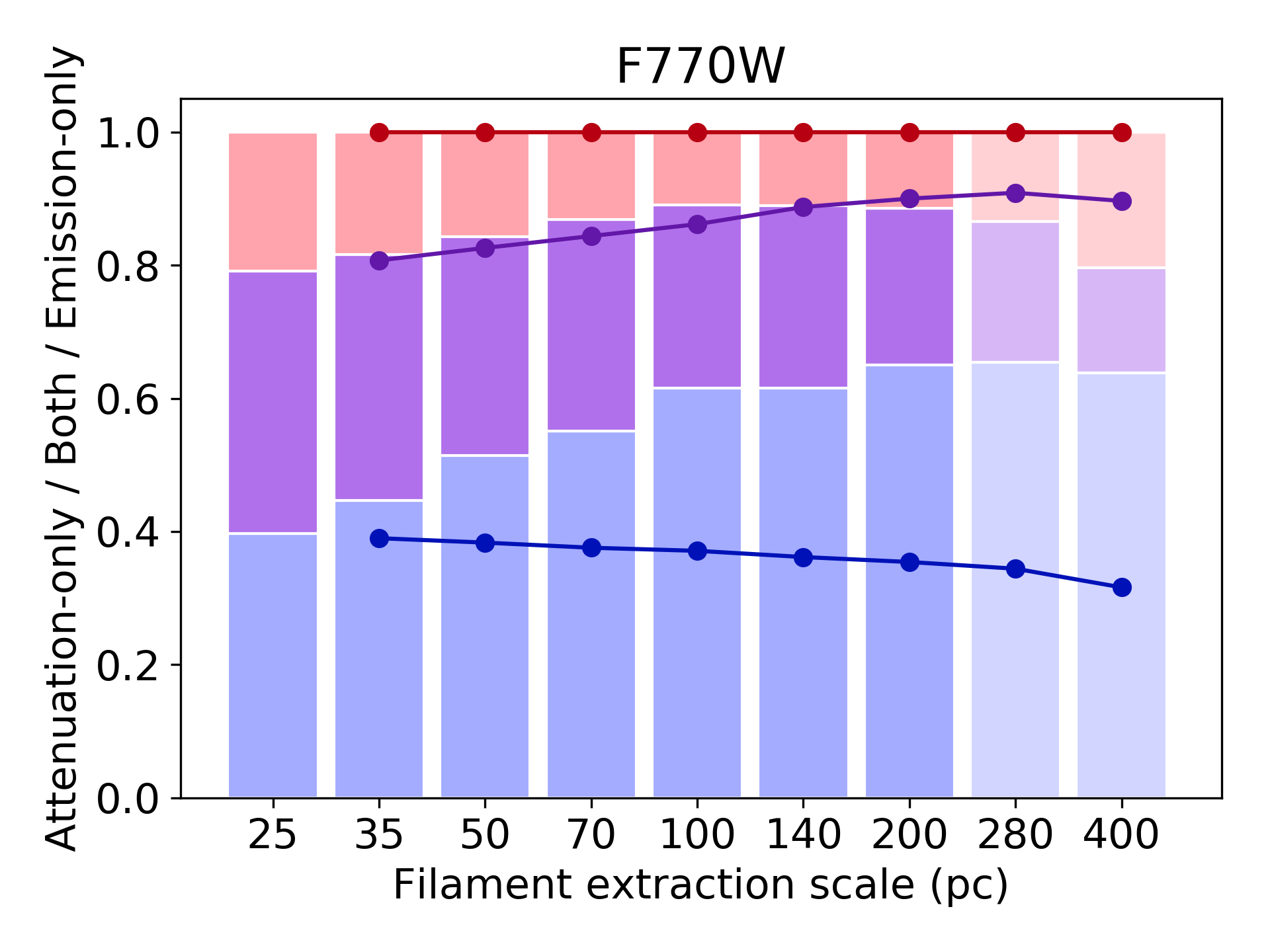}
\includegraphics[width=\linewidth]{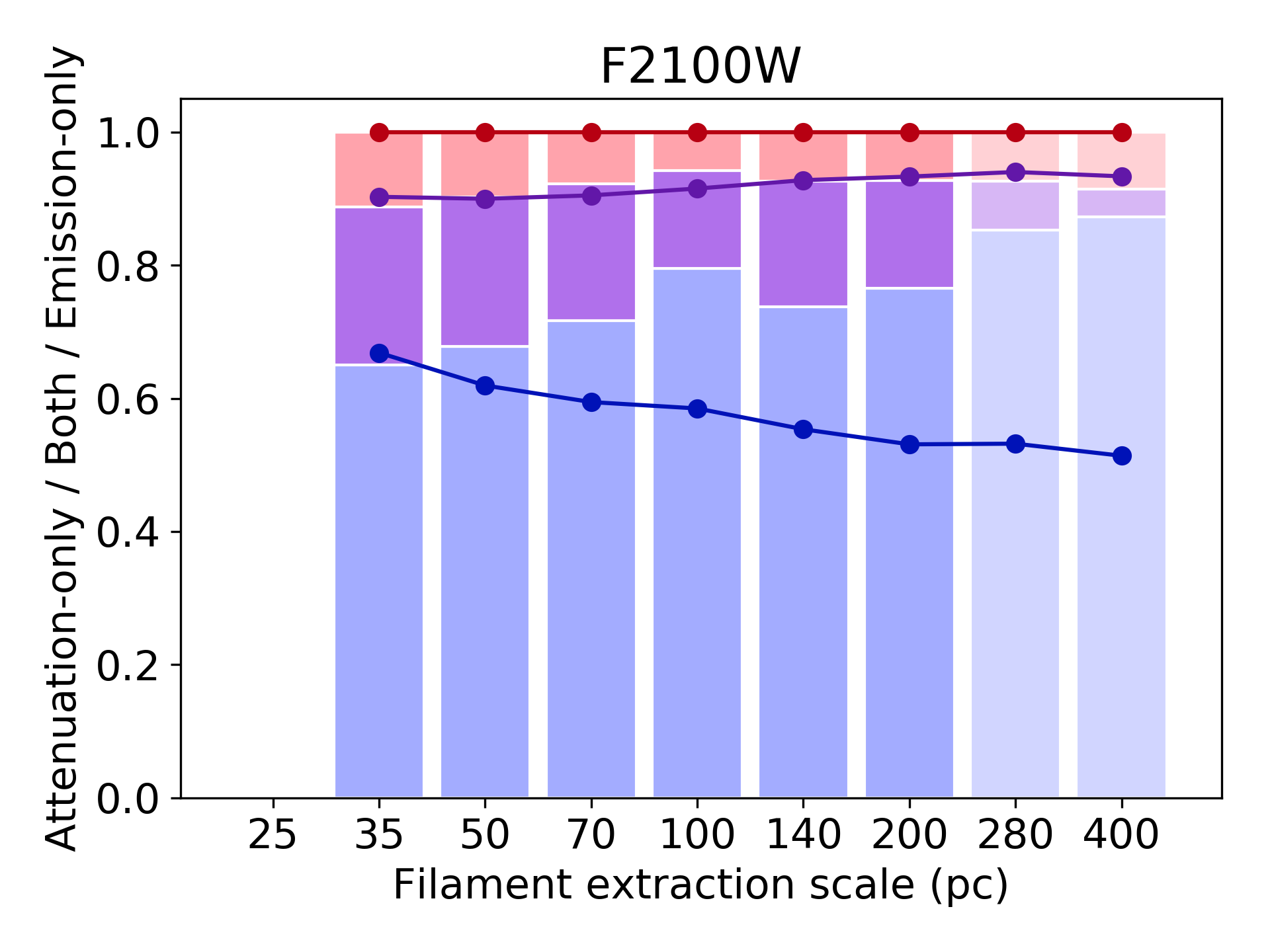}
\caption{\edit1{Filament overlap analysis for structures traced by visible attenuation and by MIR dust emission, as a function of extraction scale.} Blue markings represent sight lines with attenuation-only, magenta markings represent areas of overlap of between attenuation and emission, and red markings account for sight lines with emission-only. \textit{Top:} F770W. \textit{Bottom:} F2100W.  Results generated for cumulative masks are shown with points and sloping lines. We disregard scales $>200$\,pc (shown faded in the plot) because the filament masks become unreliable (in the case of attenuation) or redundant with smaller scale features (for emission).  The first bar of the F770W plot (25~pc) presents the measurement for masks shown in Fig.~\ref{fig:triofig}.}
\label{fig:piegraphs}
\end{figure}  

\begin{figure}
\includegraphics[width=\linewidth]{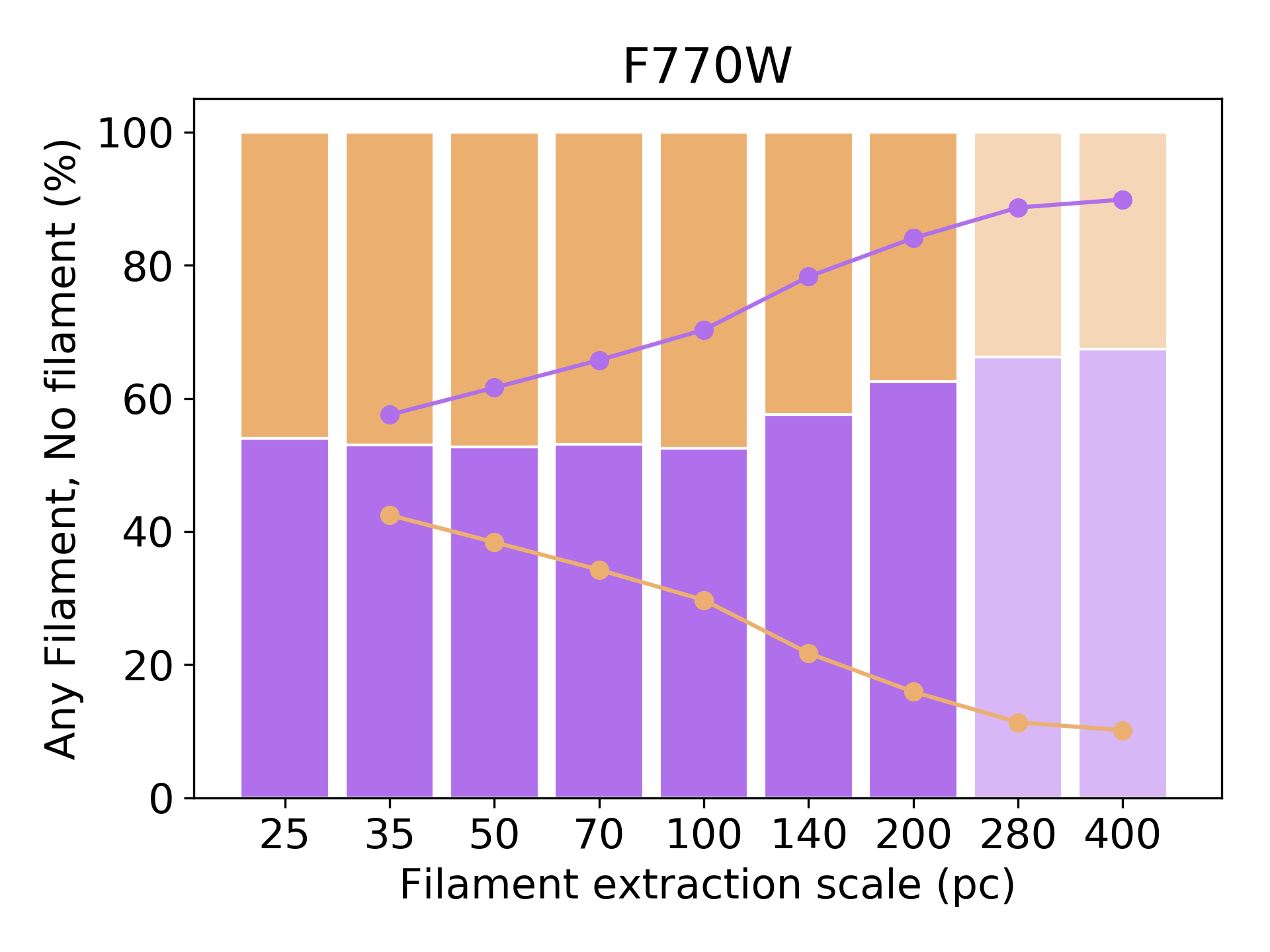}
\includegraphics[width=\linewidth]{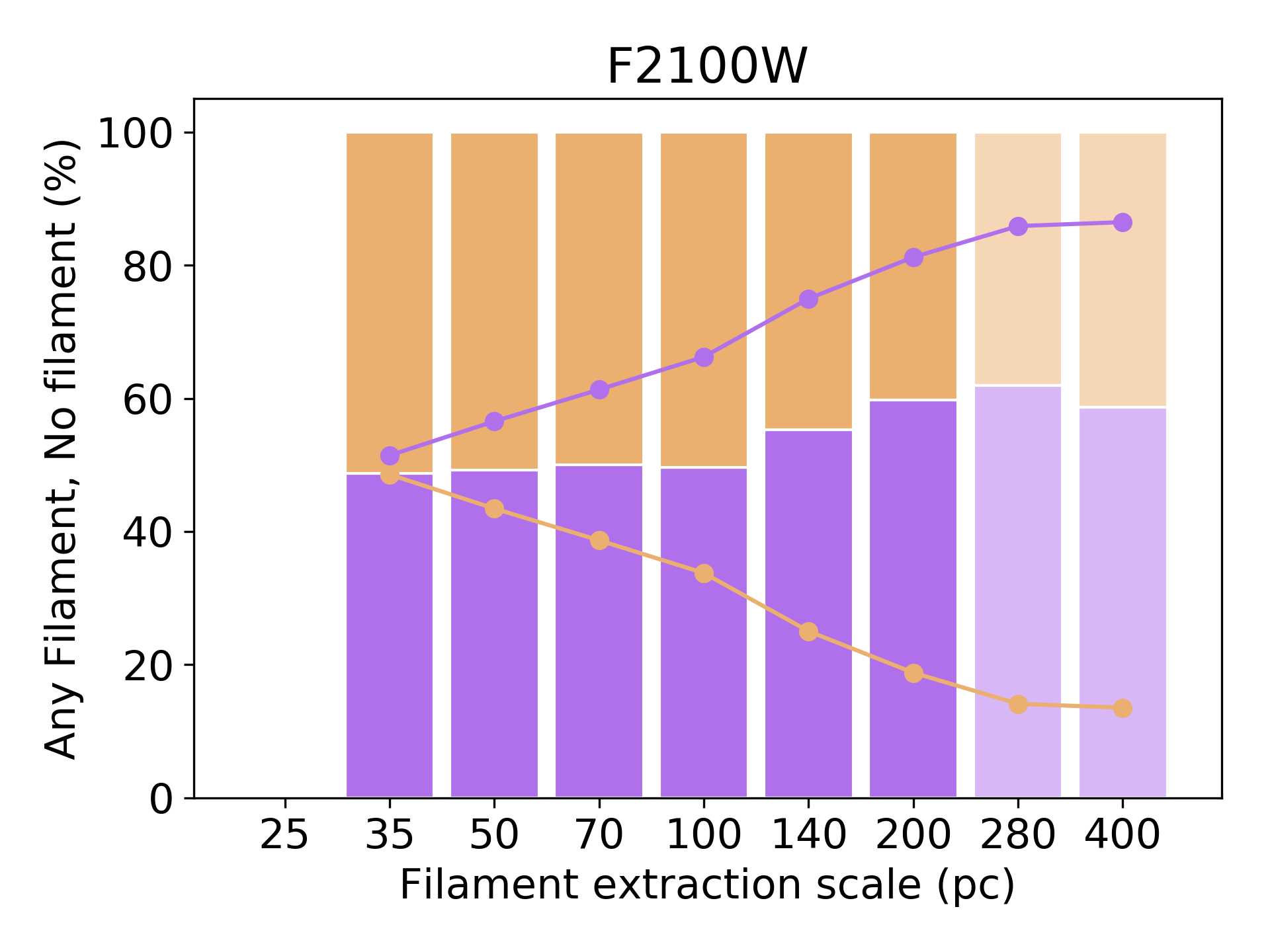}
\caption{\edit1{Areal coverage of dust filaments} \textit{Top:} For F770W, the fraction of pixels covered by a dust filament mask (emission or absorption) or left unassigned to a dust filament.  Scales we elect to exclude from the filament network (see text) are shown faded in the plot. \textit{Bottom:} Same, for F2100W.}
\label{fig:anynone}
\end{figure}  
\section{Data}
\label{sec:data}
\subsection{PHANGS-JWST imaging}\label{sec:jwstdata}
NGC\,628 is one of the initial targets observed for the PHANGS-JWST Cycle 1 Treasury project (GO 2107, PI J.\ Lee).  Our observations of NGC\,628 target the central region ($R_{gal}$\,$\lesssim$\,5\,kpc) of the star-forming disk, to overlap areas where \HST, ALMA, and MUSE data have been obtained.\footnote{Footprint maps of the \HST, ALMA, and MUSE observations are available at  https://archive.stsci.edu/hlsp/phangs-hst. Science-ready PHANGS-HST images are also available for download.}  The dataset includes imaging with NIRCam (F200W, F300M, F335M and F360M) and MIRI (F770W, F1000W, F1130W and F2100W). Photospheric emission from resolved stellar populations is \edit1{a} major component in the four NIRCam bands (excepting F335M which primarily probes the 3.3\micron~PAH feature), whereas MIRI 
traces the ISM (both PAHs and hot, small dust grains).  Resolution varies 
from 0.066 to 0.12\arcsec (NIRCam) and 0.25 to 0.67\arcsec (MIRI).  For MIRI, this corresponds to 12--32 pc at the distance of NGC\,628. A detailed description of the PHANGS-JWST observations and data reduction is presented by \cite{LEE_PHANGSJWST}. 
Here, we focus on MIRI imaging, deferring analysis of filamentary structure seen in the 3.3\micron~PAH feature to future work in anticipation of improved astrometric alignment among the PHANGS-JWST NIRCam and MIRI imaging. 
Of particular importance to the analysis in Sec.~\ref{sec:fluxfrac} is the sky background adopted for the MIRI data.  We use background-corrected images which have been tied to the sky level measured in wide-field Spitzer and WISE archival imaging, as described in the Appendix of \citet{LEROY1_PHANGSJWST}. The MIRI background levels are currently uncertain by $\pm$0.1\,MJy\,sr$^{-1}$. 

\subsection{PHANGS-HST imaging}\label{sec:HSTdata}
The \HST NUV-U-B-V-I (F275W, F336W, F435W, F555W, F814W) observations of the central NGC\,628 disk we use were obtained by LEGUS (GO 13364, \citealt{Calzetti+2015}) using WFC3/UVIS for NUV and U, and by R. Chandar (GO 10402) using ACS/WFC for B, V and I. All data were reprocessed by PHANGS-HST. Full details are given in \citep{PHANGS-Hst}. The ACS/WFC B-band images we use to identify dust lanes in attenuation have resolution ($\sim$\,0.09$\arcsec$, 4.3 pc), approximately 2.5$\times$ finer than MIRI F770W. 

\subsection{MUSE \HII region catalog}\label{sec:MUSEdata}
We use the nebular catalog of \HII regions derived from the integral field unit (IFU) spectroscopy of the PHANGS-MUSE survey \citet{Emsellem+2022}.  
For NGC\,628, the "convolved, optimized" resolution in PHANGS-MUSE DR 2.2 is $0.92\arcsec$, corresponding to a spatial resolution of $\SI{44}{\parsec}$. \citet{Santoro+2022} and \citet{GROVES_HIICAT} used PHANGS-MUSE data to create a catalog of \HII regions and provide fluxes corrected for Milky Way and internal extinction. Only star-forming regions classified 
using the BPT diagram \citep{Baldwin+1981} are retained in our analysis.

\subsection{\textit{HST} stellar association and cluster catalogs}
PHANGS-HST resolved stellar photometry has been used to identify and characterize stellar associations as summarized in \citep{PHANGS-Hst} and described in detail by \citet{LARSON_MSA}.  Stellar clusters in NGC\,628 have been studied by \citet{Thilker2022a}. 
PHANGS-HST catalogs are publicly available\footnote{https://archive.stsci.edu/hlsp/phangs-cat}. 
For both associations and clusters, fluxes for the five available \textit{HST} bands were measured (using upper-limits in non-detected photometric bands) and then age, mass and reddening were estimated\footnote{Degeneracy between age and reddening is apparent for a subset of objects, and could be more relevant in our dust filaments.} via fitting of observed SEDs \citep{Turner2021} to solar metallicity stellar population models 
using \textsc{cigale} \citep{Boquien+2019}.

We use the PHANGS-HST associations catalog based on local over-densities of NUV point-like detections at a scale of 32 pc for our analysis. 
The stellar associations have ages ranging up to $\sim$10$^2$ Myr. We use a subset selected to have age less than five Myr in order to limit the population to the most recent star formation activity.   Clusters for our analysis were also selected with the same upper limit on age.

\section{Dust filament analysis}
\label{sec:analysis}

\subsection{Filament extraction}
\label{sec:filextraction}
We identify dust filaments using \textsc{FilFinder} \citep{KochRosolowsky2015}. This code applies an adaptive thresholding algorithm and graph-based medial skeleton analysis to isolate and then characterize filaments. Thresholding is conducted over local neighborhoods, allowing for the extraction of structure over a large dynamic range both in intensity and spatial scale (the latter when the code is run multiple times with different parameter choices). The potential effect of bright sources interspersed in the web of filaments is mitigated by an arctan intensity transform before thresholding. We use a slightly modified version of \textsc{FilFinder} in which the arctan transformed image is convolved to the filament extraction scale before each run. In the current analysis we only utilize the filament masks produced by the code, leaving \textsc{FilFinder}'s skeleton analysis capabilities for future work.   

We apply \textsc{FilFinder} independently to the background-corrected image in each MIRI band and to a pre-processed version of the B-band \HST image (in which attenuation features are most evident compared to NUV, U, V, and I). 

For pre-processing the \HST image, multi-scale median filtering is used to suppress peak-like features over a range in scale, while 
retaining small scale dips (concave areas).  Our specific filtering method follows from 
\citet{Hoversten2011}.  At each pixel 
the output value is assigned to be that location’s minimum in a stack of circular median filtered images. Filter kernel diameters are taken from a ladder of physical scales (starting at the resolution limit and proceeding up to 32 pc).  The result is that confusion by bright stars and stellar clusters is greatly reduced, emphasizing the dust lane structures. Fig.~\ref{fig:triofig} (left) and Fig.~\ref{fig:dustmotes} (left) show the pre-processed image.  \textsc{FilFinder} nominally operates by finding positive filamentary features above the background.  \edit1{Therefore,} before passing the pre-processed B-band image to \textsc{FilFinder}, we invert the sense of the intensity (subtracting the image from a constant value equal to the maximum in the field of view)\edit1{, so that dust attenuation filaments appear as the positive features expected by \textsc{FilFinder}.}

For both attenuation and emission, we use \textsc{FilFinder} to identify potential filaments with narrow dimension (width) 
starting at 25~pc then stepping by factors of $\sqrt{2}$ (0.15 dex) up to 400~pc (25, 35, 50, 70, 100, 140, 200, 280, 400~pc). The minimum scale of 25~pc 
corresponds to approximately twice the PSF FWHM of MIRI F770W.  For F2100W analysis, we begin at the 35~pc scale due to the 
larger F2100W PSF. \textsc{FilFinder} parameters are set as follows: size\_thresh\,=\,6$\pi$$w^2$, adapt\_thresh\,=\,2$w$, glob\_thresh\,=\,2$\sigma$ above sky level, smooth\_size\,=\,0.5$w$, fill\_hole\_size\,=\,0.5$w^2$, where $w$ is the extraction scale (in pixels) and $\sigma$ is the standard deviation noise level expected in the arctan transformed, convolved images. \edit1
{We were guided in these decisions by the discussion of methodological testing included in \cite{KochRosolowsky2015} and the online \textsc{FilFinder} code tutorial / test suite. The specific choice of parameters has minimal influence on the resulting filament masks but this will be quantified fully in a follow-up paper analyzing the entire set of PHANGS-JWST targets.  One particular planned aspect of expanded testing is bench-marking in the multi-scale extraction regime. Such a future study will also ideally include JWST-based comparison of \textsc{FilFinder} to other codes for filament extraction \citep[e.g.][]{Sousbie2011,Menshchikov2021,Menshchikov2021b,Carriere2022,Alina2022}.  Note however that \textsc{FilFinder} has been successfully utilized for astrophysical analysis in many works and in some instances compared to other codes \citep[e.g.][]{Green2017}.}

In addition to the filament masks generated for extraction at specific scales, we also produce masks 
representing the union of filaments detected cumulatively at different scales.  For these cumulative masks, we sum the individual masks (for scales less than or equal to the current scale) and then flatten the result, such that it has a value of one anywhere a constituent scale contributes filament coverage.  The cumulative summed mask before flattening is also retained, as it highlights the multi-scale nature of the filament network.  Appendix~\ref{sec:appendix} presents individual scale and cumulative multi-scale masks for F770W, F2100W, and \HST B-band as a Figure Set. 

The wide range of scales initially allowed for filament extraction is exploratory.  In the second half of 
Sec.~\ref{sec:twoviews}), we argue that all emission and attenuation filament masks for scales $>200$\,pc not be used, although we do include them in plots allowing for the reader to make their own \edit1{judgment}. 

\subsection{DFN characterization}
\label{sec:filcharacterization}
For Sec.~\ref{sec:twoviews}, we measure filament mask overlap
categorizing each pixel as belonging to one of four classes: (1) attenuation filament only, (2) emission filament only, (3) both attenuation and emission filaments, or (4) no filamentary features detected.   Note that the sight line fractions we report in Sec.~\ref{sec:twoviews} for classes (1)-(3) are normalized to the total count of pixels with any detected filamentary feature, rather than the total count of pixels in the image.   It is beyond the scope of the current study to quantify the dependence of pixel classes on the depth of the imaging -- however, given the pervasive character of the detected filament network, we suggest that at least the F770W, F1000W, and F1130W observations are sensitive enough to make this a moot point.  Our F2100W data is about 2$\times$ less sensitive in absolute terms of limiting surface brightness, $\sigma_\mathrm{I}$ [MJy\,sr$^{-1}$], and also suffers from a similar loss in resolution compared to F770W.  These factors likely contribute to loss of some smaller scale filamentary structure in the F2100W imaging.

Sec.~\ref{sec:fluxfrac} (flux fractions) requires local estimation of the diffuse emission 
to obtain background-subtracted filament flux.  \edit1{Methods of background determination are highly varied (e.g. circular annuli, region-hugging annuli, background surface determination) and generally selected on the basis of the specific use case. We employ} 
the \textsc{photutils} \citep{Bradley2022} \texttt{background2D} code, supplying filament masks to indicate which pixels the procedure should ignore.  A mesh of bins (each having size two-thirds the filament extraction scale) is defined, and the mode of unmasked pixels is determined in each bin.  These mesh modes are median filtered with a 3$\times$3 boxcar (ignoring bins with too many pixels masked as within a filament) and the resulting values are interpolated across the image grid.  \edit1{Finally, the \texttt{background2D} output is  subtracted} from the input image and non-filament pixels are set to zero.  Integrating this result gives the background subtracted flux of the filament structures, which is then divided by the total flux in the MIRI footprint to obtain flux fractions. 
\section{Results}
\label{sec:results}
Figure~\ref{fig:triofig} illustrates the dust lanes seen as deficits of visible light (left) and dust emission filaments detected in MIR emission (right).  Plotted for a single extraction scale (25~pc), the central panel emphasizes the detailed coincidence between these two tracers of dust (blue = attenuation, red = emission) in the interstellar medium, with magenta indicating overlap.
It is clear that the dust filament network (DFN) occupies a large fraction of sight lines and contributes a significant fraction of the MIR luminosity \edit1{of NGC\,628}.  Knots of emission from star-forming regions are generally distributed 
throughout the filaments.  In this section, we quantify each of these statements. 

Here we show results only for F770W and F2100W, as the filament masks and measured quantities based on F1000W and F1130W are consistent with F770W. 
Any notable differences are discussed. 
\begin{figure*}
\includegraphics[width=\linewidth]{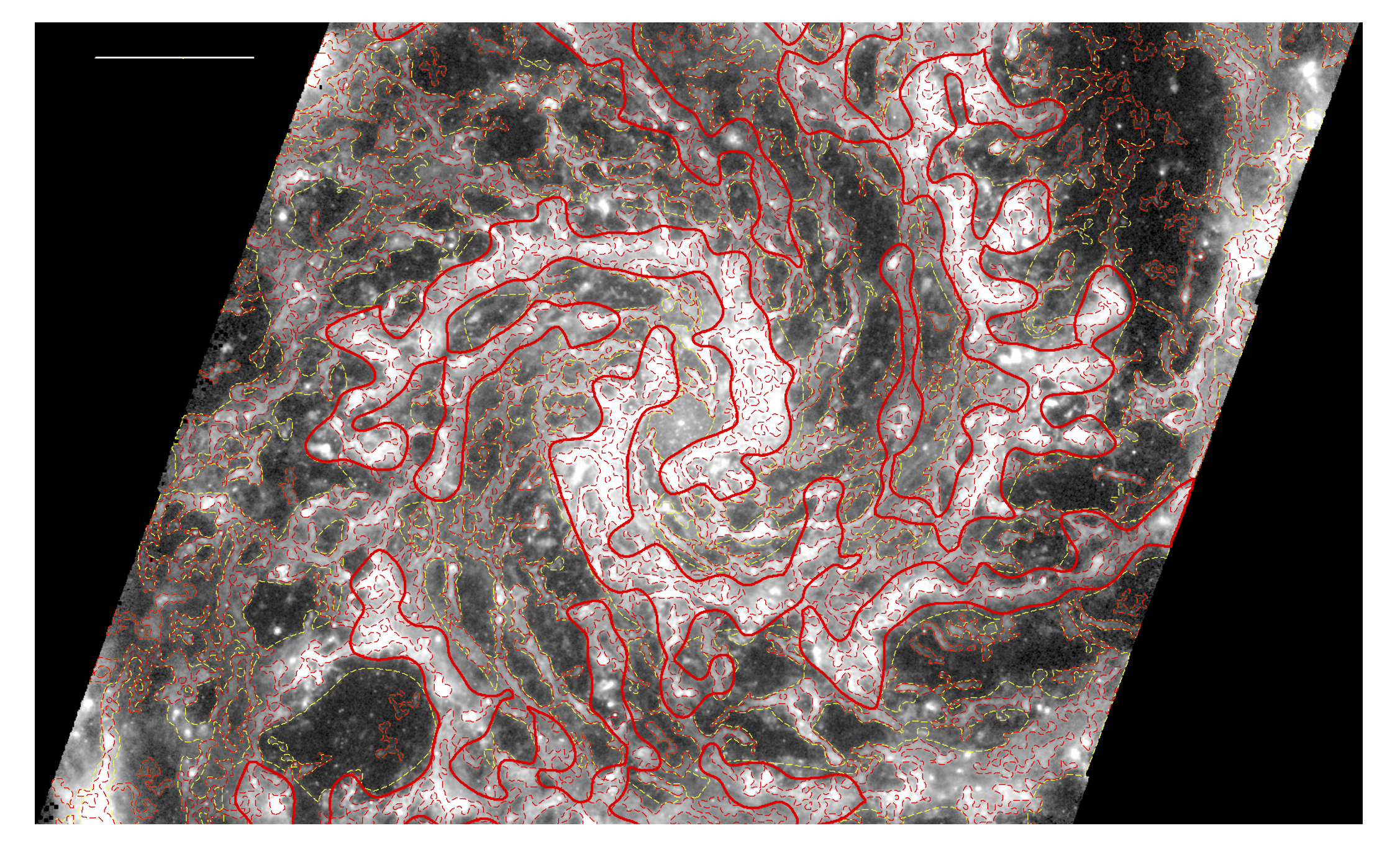}
\includegraphics[width=\linewidth]{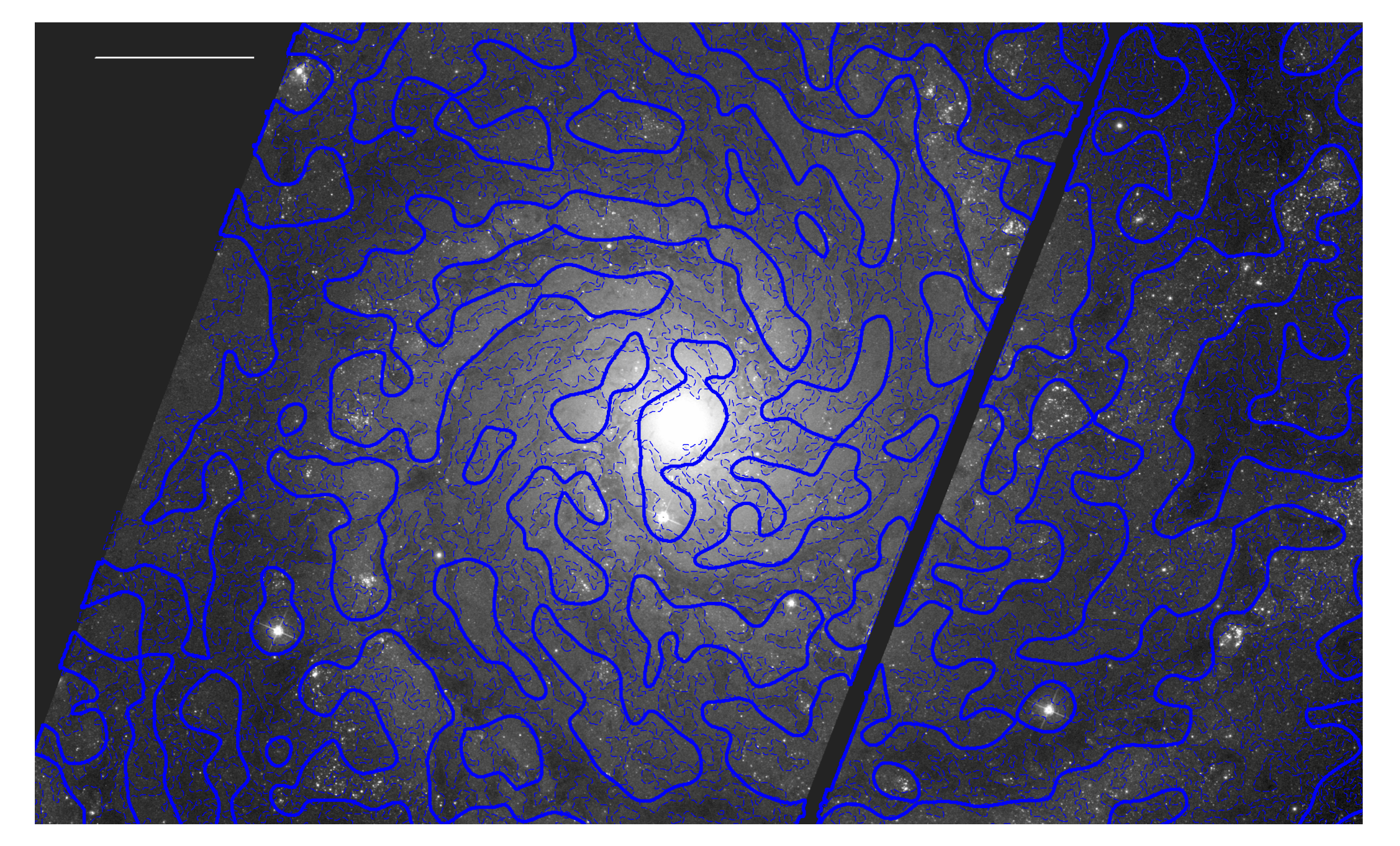}
\caption{\edit1{Contrasting filamentary dust structures in NGC\,628 as seen in emission and attenuation, with emphasis on features identified at varied filament extraction scale.} \textit{Top:} \JWST F770W image. Dust emission filament mask boundaries are shown for: 25~pc dashed, thin red; 200~pc, thick red; and cumulative 400~pc, dashed thin yellow. \textit{Bottom:} B-band F435W \HST image, with attenuation filament mask coverage of 25~pc dashed, thin blue; 200~pc, thick blue.  The scale bar in each panel is 1~kpc in length\edit1{, and the images are oriented with North up and East left.}}
\label{fig:showscales}
\end{figure*}  

\begin{figure}
\includegraphics[width=\linewidth]{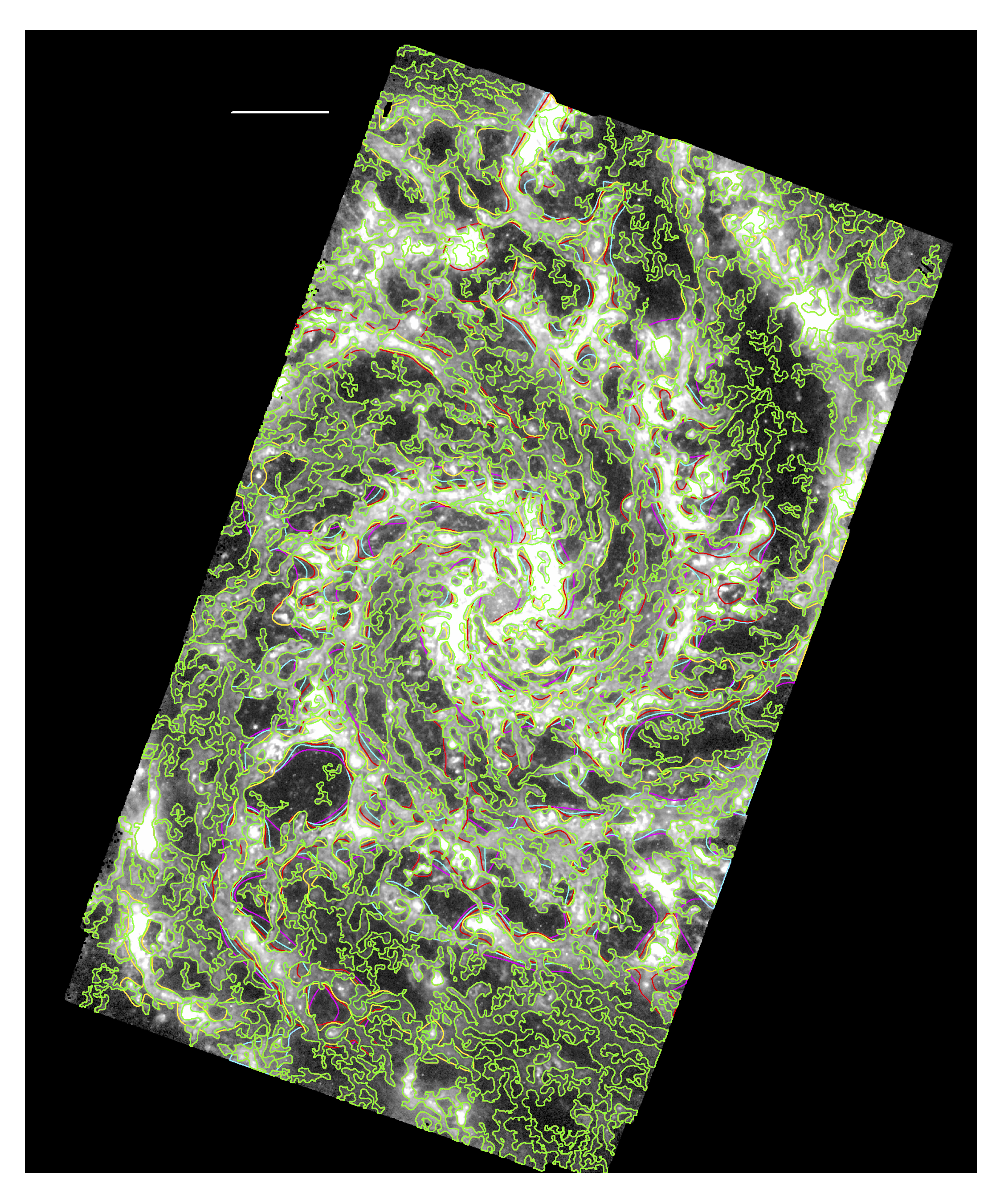}
\caption{\edit1{Comparison of cumulative filament masks up to varied maximum scale.} \JWST F770W image of NGC\,628, overplotted with the contours showing the coverage of multi-scale masks. The maximum scale is as follows: green, 100~pc; yellow, 140~pc; red, 200~pc; cyan, 280~pc; magenta, 400~pc. We adopt 200~pc as a preferred value. The scale bar corresponds to 1~kpc\edit1{, and the images are oriented with North up and East left.}}
\label{fig:comparecumulativescales}
\end{figure}  

\begin{figure*}
\includegraphics[width=\linewidth]{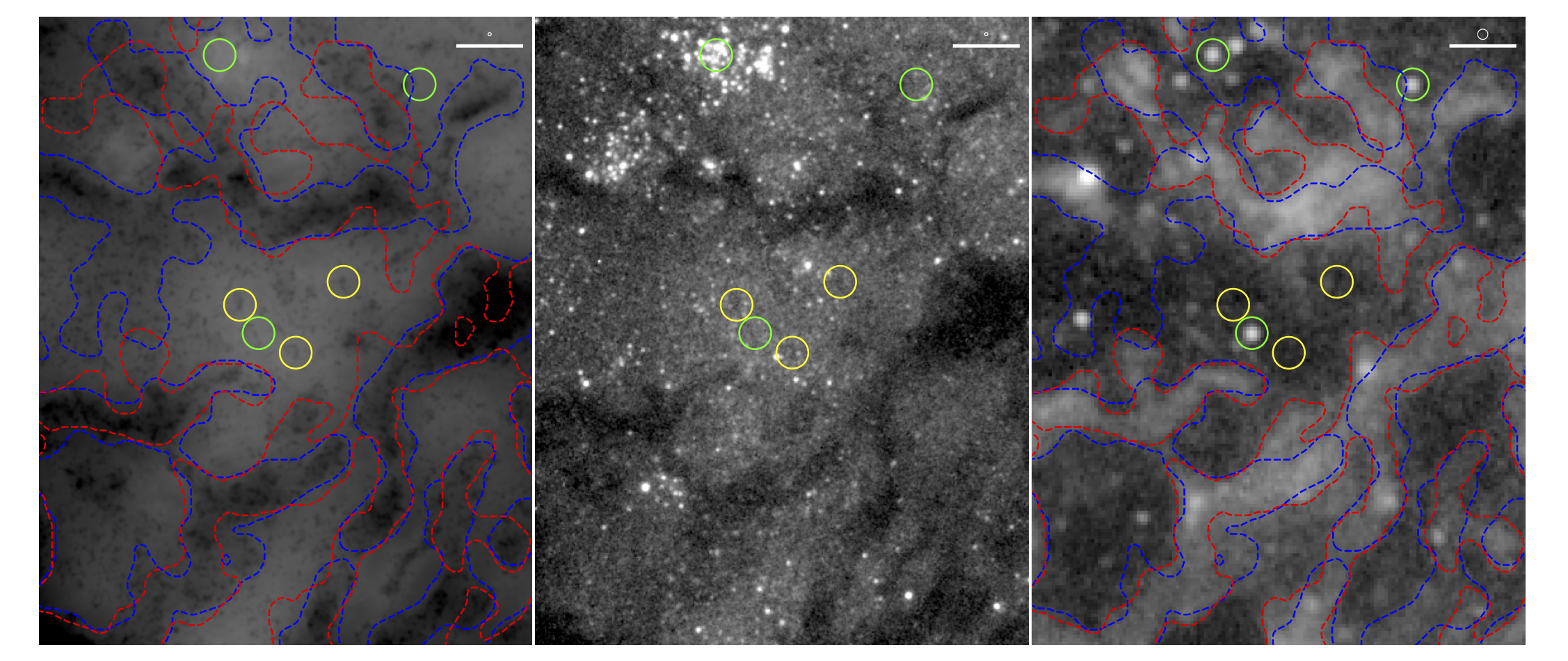}
\caption{\edit1{Dust motes and point-source MIR contamination from extremely dusty stars.} \textit{Left:} Subsection of our pre-processed B-band image, overlaid with contours corresponding to the filament masks generated at 25~pc resolution in emission, red, and absorption, blue. \textit{Center:} Unprocessed B-band image \textit{Right:} \JWST F1000W image, also with the filament mask contours shown.  Yellow circles surround three example dust motes.  There are many more in the field shown, left unmarked. Green circles surround three example candidate dusty stars, others are not marked. Circles are $1\arcsec$ in diameter, equivalent to 48~pc at the distance of NGC\,628\edit1{, and the images are oriented with North up and East left.}}
\label{fig:dustmotes}
\end{figure*}  

\begin{figure}
\includegraphics[width=\linewidth]{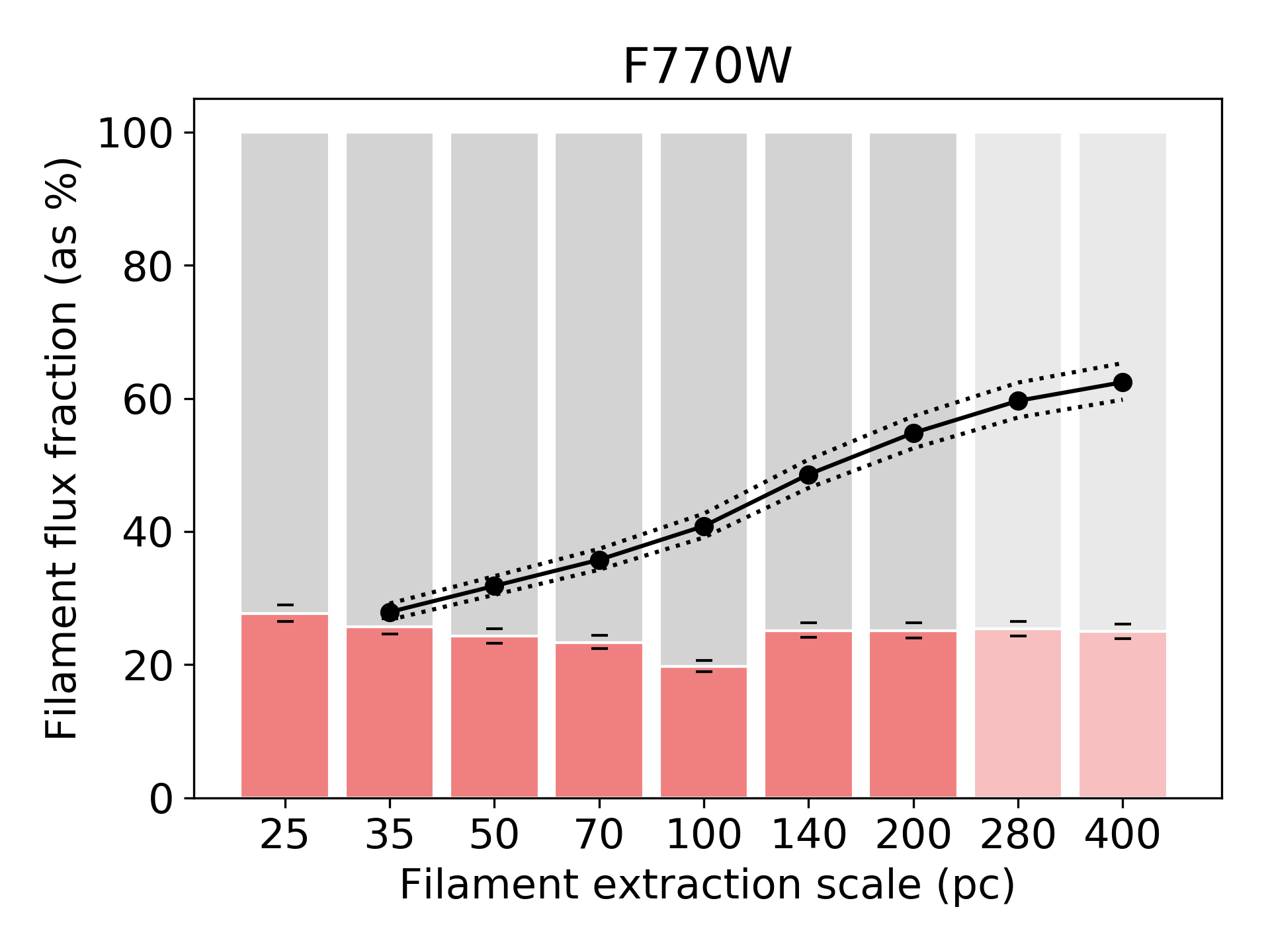}
\includegraphics[width=\linewidth]{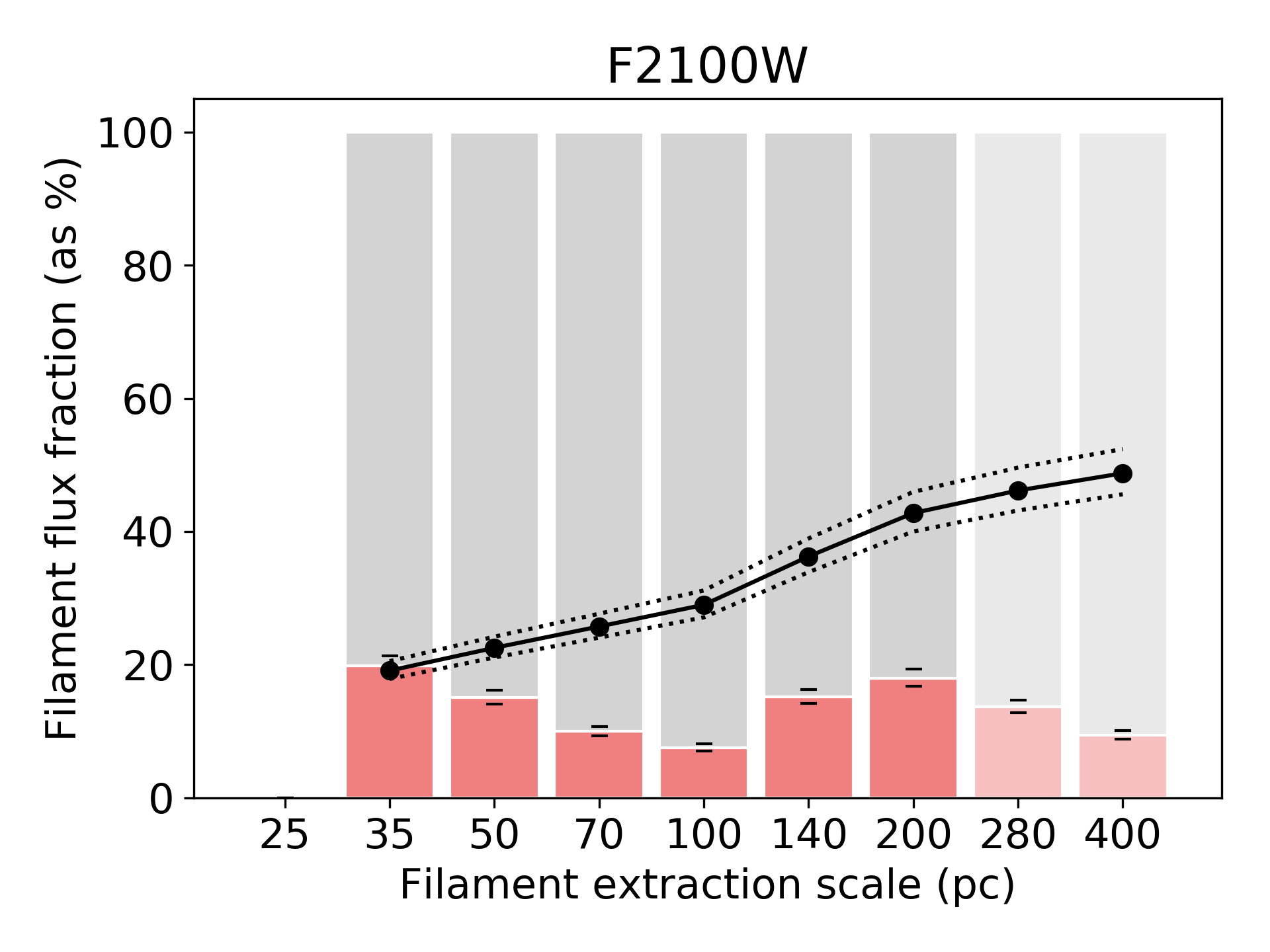}
\caption{\edit1{Measurements of the fraction of total flux contained within the dust filament network} \textit{Top:} Filament flux fraction in F770W expressed as percentage, with individual filament extraction scales plotted as pink bars, and cumulative multi-scale results as solid lines and points.  The faded region is shown for completeness, although we adopt a maximum cumulative scale of 200~pc. Horizontal dashes on each bar indicate the one-sigma uncertainty of the flux fraction measurements, obtained by perturbing the sky level in accord with the post-pipeline calibration described by \citet{LEROY1_PHANGSJWST}, \citet{LEE_PHANGSJWST}. Dotted lines show the one-sigma range due to uncertainty of cumulative flux fractions. \textit{Bottom:} Same, but for F2100W.}
\label{fig:fluxfrac}
\end{figure}  

\begin{figure*}
\includegraphics[width=0.5\linewidth]{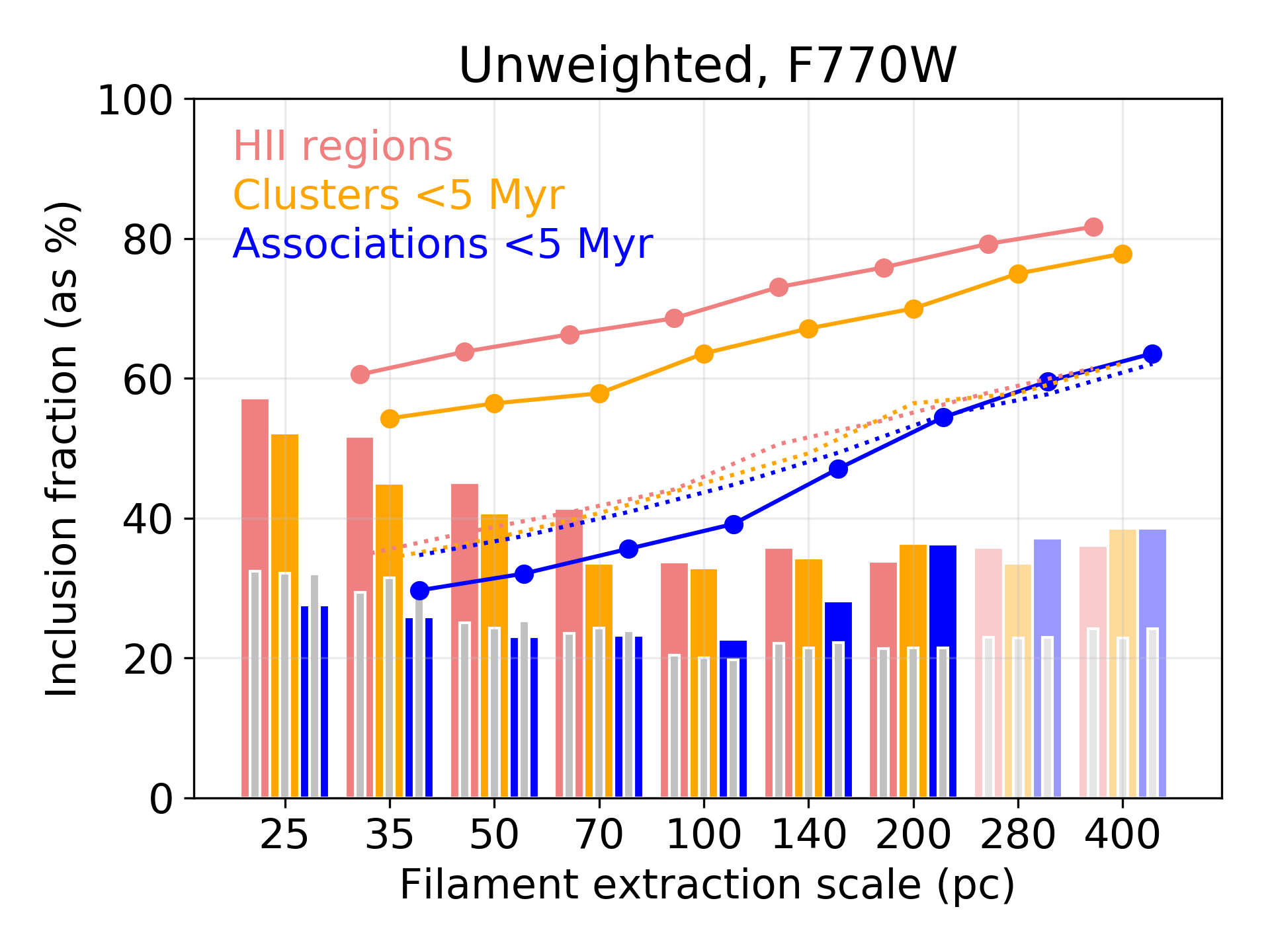}
\includegraphics[width=0.5\linewidth]{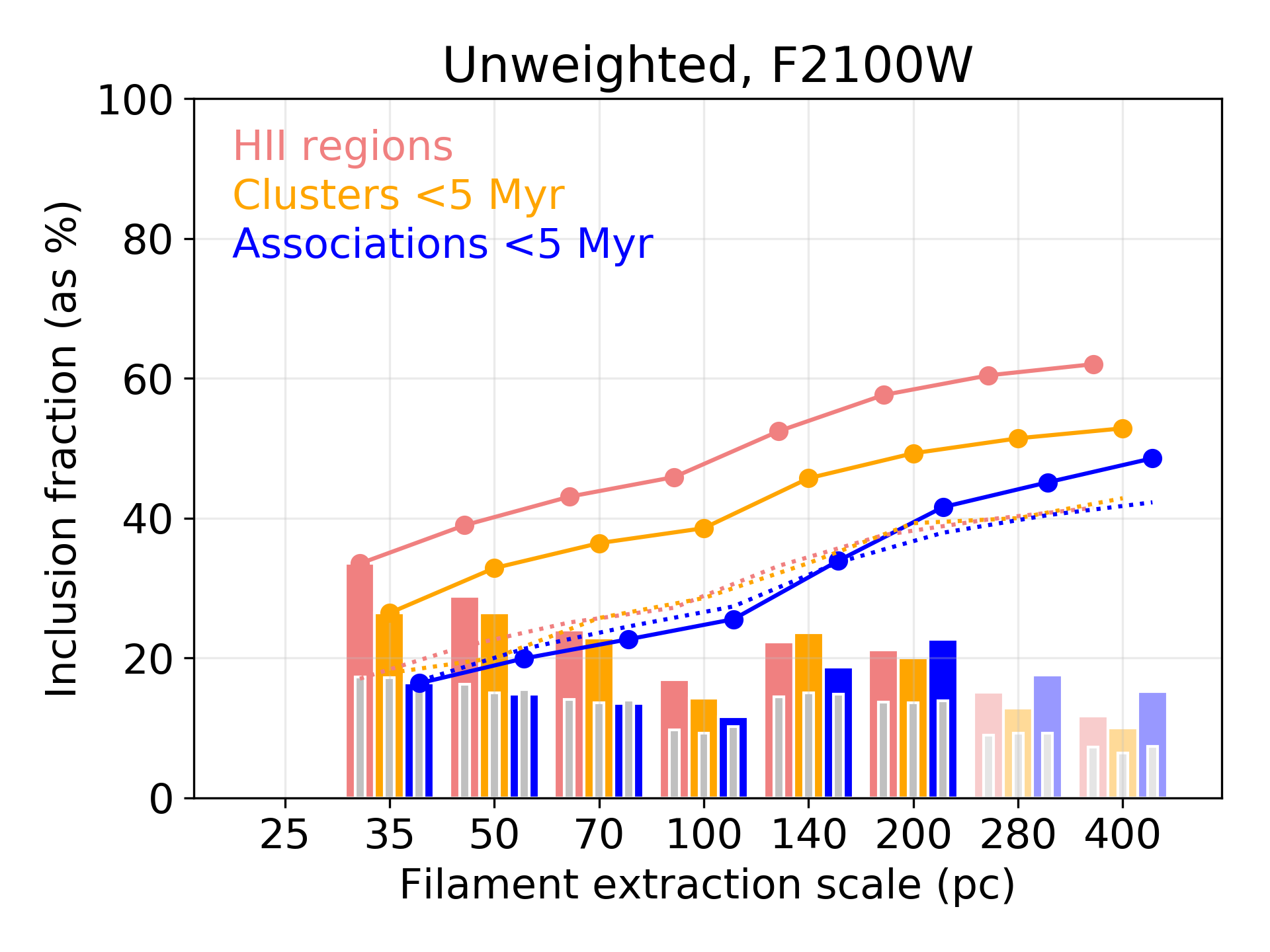}
\includegraphics[width=0.5\linewidth]{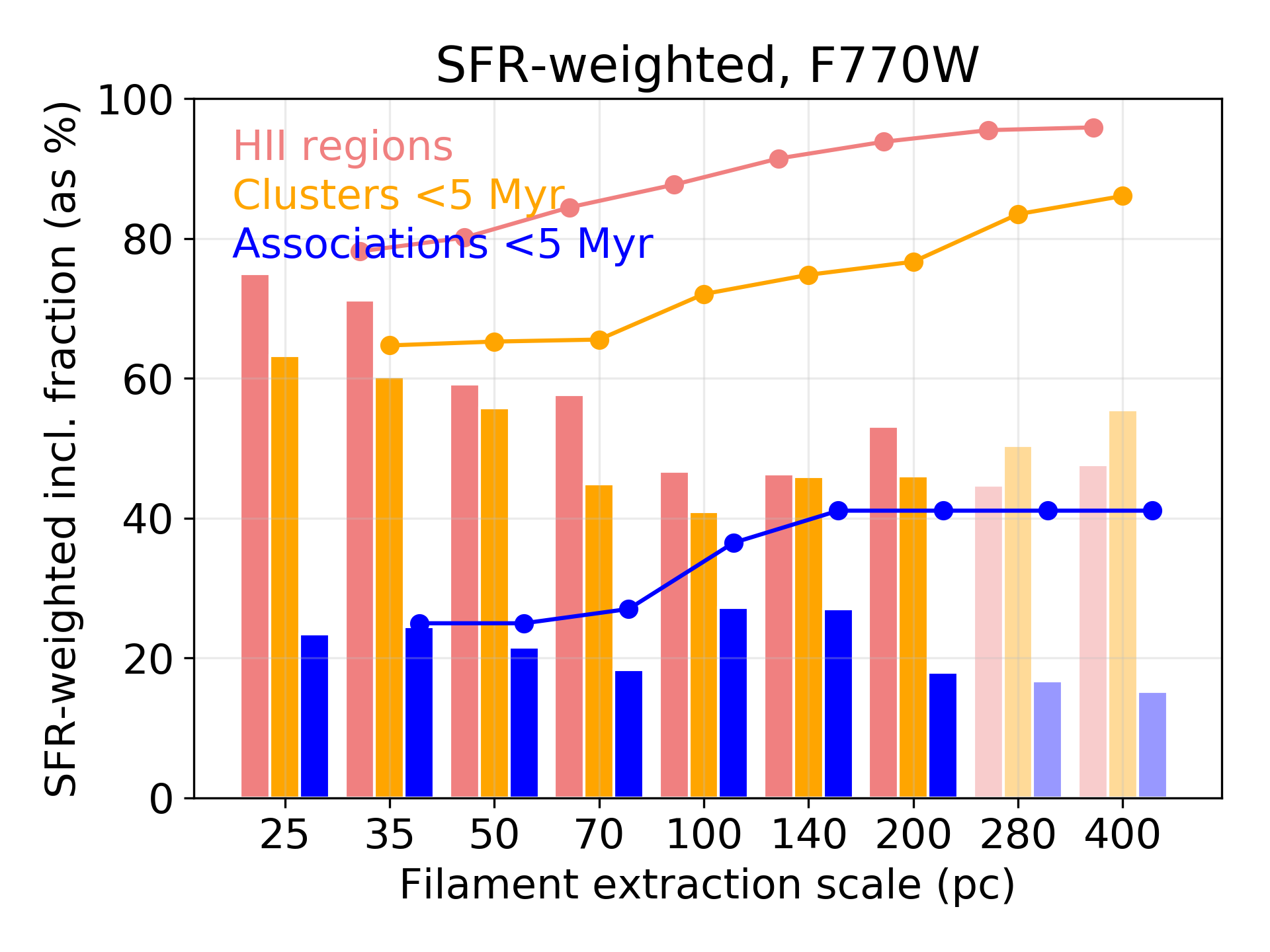}
\includegraphics[width=0.5\linewidth]{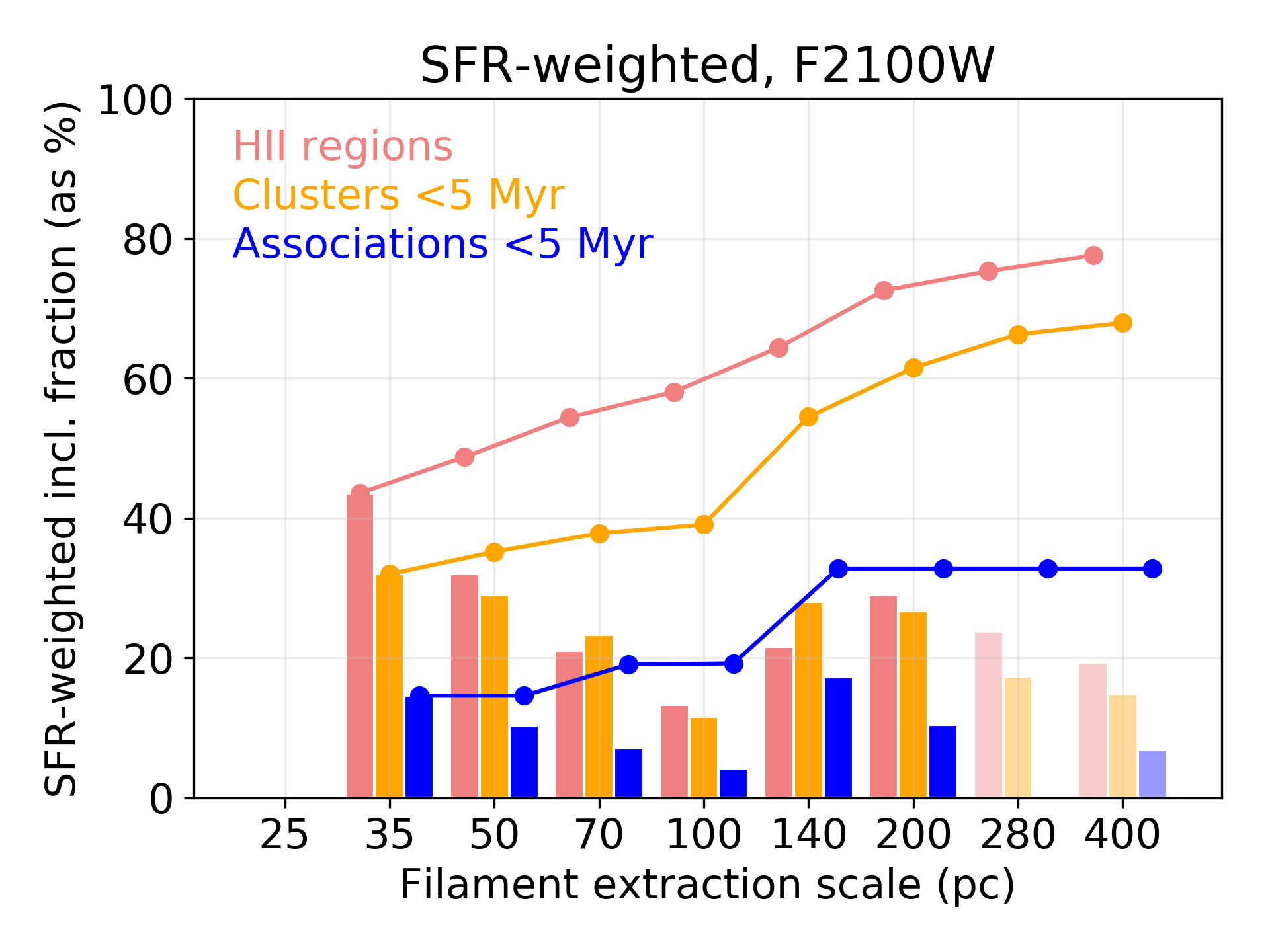}
\caption{\edit1{Testing the link of the dust filament network to current star formation.} \textit{Top:} Filament inclusion fractions for various star formation event tracers (\HII regions, young clusters, young multi-scale associations) versus the dust emission filament extraction scale. Bars indicate results for each individual extraction scale, whereas lines and points indicate the cumulative filament mask measurements. Inclusion fractions for randomly distributed points are shown with inset bars and dotted lines. \textit{Bottom:} SFR-weighted inclusion fractions.   }
\label{fig:piegraph_InclFrac_vs_scale}
\end{figure*}  

\subsection{Contrasting views of the dust filament network}
\label{sec:twoviews}

We start by highlighting the consistency between filamentary attenuation (dust lane) features and web-like MIR dust emission   
to illustrate the potential of using \HST-detected features as a 
high resolution proxy for the dense dusty ISM morphology (and perhaps even molecular gas\footnote{The association between the DFN structures and $^{12}$CO(2-1)-traced molecular gas will be investigated in a future study.}).  

Fig.~\ref{fig:piegraphs} shows the results of filament overlap analysis.  
For the PAH-dominated bands (F770W, F1130W) and 10$\micron$ thermal emission traced by F1000W, in the top panel of Fig.~\ref{fig:piegraphs}, we find that the percentage of sight lines in common between visible attenuation and MIR emission filaments is nearly 40\%, at 25~pc and declines smoothly with increasing scale (filament width). This decline is due to more area becoming traced by attenuation only for larger individual filament extraction scales. 
The percentage of emission-only filament sight lines declines a small amount from 25~pc to 200~pc. Overlap statistics generated on the basis of cumulative multi-scale masks show a different picture, in which the percentage of attenuation plus emission sight lines grows with scale from just over 40\% at 35~pc to 55\% 
up to 200~pc. This 
is a consequence of different extraction scale masks picking up varied portions of the overall web-like DFN.  We return to this point further in the current subsection (see Fig.~\ref{fig:showscales} and Fig.~\ref{fig:comparecumulativescales}). 

In the bottom panel of Fig.~\ref{fig:piegraphs}, we show the overlap 
for F2100W emission filaments with 
the dust lanes from \HST.  At 21$\micron$ the percentage 
is 24\% for 35~pc individual extraction scale, substantially less than for the three other bands,  declining to $\sim$16\% for 200~pc.  For all separate scales, attenuation-only sight lines amount to more than 65\% and emission-only $\lesssim$10\%.  Given that the attenuation filament mask remains constant \edit1{in comparison to} F770W and F2100W, this could suggest that either we are sensitivity limited for F2100W or the filamentary emission detected in 21$\micron$ imaging is less consistently recovered into coherent structures.  The latter interpretation is supported by mask inspection in the figures of Appendix~\ref{sec:appendix}, and by the fact that the morphology of the F2100W image is more dominated by compact sources (e.g. IR-bright star-forming regions, see \citealt{HASSANI_PHANGSJWST}) than F770W, F1000W, and F1130W. This serves as a reminder that we are tracing warm (140~K) dust at 21$\micron$, whereas the dust attenuation provides a more complete inventory with respect to dust over a wide range of [cooler] temperatures. Nevertheless, the overlap with the 200~pc cumulative multi-scale mask 
is 40\% (only $\sim$1/3 less than for F770W).

In summary, the 40\% level of pixel-by-pixel agreement of the attenuation and emission filament masks at 25~pc 
and the trend for the greatest agreement on the smallest scales \edit1{suggest} that the 
correspondence 
may be even tighter if smaller physical scales are probed, such as in Local Group galaxies.  We stress that, although 40\% overlap may sound low, inspection of Fig.~\ref{fig:triofig} shows that the detailed morphology (e.g. extent, shape) of filaments that are detected in attenuation and dust emission is frequently rather well-matched. 
We also note that regions of the filament network only found as dust emission are expected 
due to line-of-sight effects: some filaments 
will be positioned on the `back' side of the face-on galaxy disk, and could account for the majority of `emission-only' areas.  
\edit1{The attenuation-only regions could be explained in multiple ways. 
As noted above, the dust temperature distribution, but also grain size distribution and PAH fraction \citep{CHASTENET1_PHANGSJWST,CHASTENET2_PHANGSJWST,EGOROV_PHANGSJWST,DALE_PHANGSJWST}, can influence the degree to which the MIRI bands are a complete tracer of the dust distribution.} Some attenuation-only regions also lie in false-negatives for MIR dust emission on the scale of interest, or lie in relatively MIR-faint, intermediate surface brightness regions of the visible galaxy disk. False negatives occur when a MIRI filament structure has a width that is somewhat different than the visible attenuation feature, or when scale-matched emission in the region is biased against detection at the scale of interest (e.g. a filament centered between brighter neighboring emission features spaced by about the width of the \textsc{FilFinder} adaptive thresholding box).  

\edit1{One might wonder if segmenting the galaxy into filaments is actually necessary, or if it might be just as useful to characterize the level of correlation between emission and attenuation with a pixel-level scatter plot between the two quantities.  As described in Sec.\ref{sec:summary}, such analysis will be included in future work, but even then the estimation of attenuation will require division into filament and non-filament regions unless the properties of the local stellar population behind (or mixed with) the dust is otherwise known.  Furthermore, the focus of the present paper on filament morphology/overlap and the association with star formation activity requires segmentation.} 

Figure~\ref{fig:anynone} shows the division of sight lines between those associated to a dust filament (here, either attenuation or emission) and those that do not lie in a filament for the particular extraction scale.  The plots show very little change between bands. We find that approximately 50--55\% of sight lines are attributed to dust filaments for scales 25--100~pc with no variation due to extraction scale, then a mild increase at larger scales (to $\sim$60--65\% for 200~pc). Cumulative multi-scale mask measurements of the dust filament sight line fraction (dots and lines in Fig.~\ref{fig:anynone}) steadily rise across the 25--200~pc range, despite the constancy for individual extraction scales smaller than 140~pc.  Our filament masks are detecting a morphologically diverse dusty ISM, \edit1{with features ranging from very narrow filaments like the Milky Way's `bones' / `Giant Molecular Filaments' to lower density, more diffuse dust filaments.} Though beyond the scope of this paper, it will be important to investigate if the \edit1{latter} component is increasingly atomic-dominated compared to narrower filaments.

Figure~\ref{fig:showscales} illustrates the tendency for dust filament masks to capture different morphological features when extracted using varied scales.  In particular, we show the \JWST F770W image in the top panel and \HST B-band in the bottom panel.  Using red and blue lines, overplotted on each of the images is a contour boundary of the 25~pc mask (dashed thin line) and 200~pc mask (solid thick line), with the variety of mask corresponding to the dust detection property of each image (emission on top, attenuation on bottom). Many of the single extraction scale 25~pc dust filaments are very long, with $l>$1~kpc and have rather high aspect ratio.  However, there are also frequent cases of the 25~pc masks (dashed thin lines) only including localized substructure within larger coherent filaments left unjoined by small scale extraction. This selective property of the filament identification outcome can be seen in spiral arms (markedly less so in interarm regions) and in both emission and attenuation. Conversely, the 200~pc filament masks (solid thick lines) often entirely exclude areas with a significant population of narrow GMC-scale width filaments. We note that the 200~pc attenuation filament mask recovers continuous spiral structure more effectively than the F770W emission mask of the same scale. Additionally, Fig.~\ref{fig:showscales} demonstrates that the recovered dust attenuation features can be quite modest in terms of apparent $A(B)$.  Inspection in the peripheral areas of the \HST panel nevertheless suggests the majority of such filaments are real.  On the \JWST image, we also plot a contour representing the cumulative (up to 400~pc) emission filament mask (dashed thin yellow). The 25~pc and 200~pc filament structures alone do not include all portions of the DFN, especially in dust emission.  The yellow contour shows how using a cumulative mask addresses this issue, linking many smaller scale components.  Using cumulative masks with large upper scales can excessively broaden the extent of filaments, and hence we urge caution in the choice of the upper limit on multi-scale integration.  

Fig.~\ref{fig:comparecumulativescales} provides empirical justification for maximum scale on the basis of emission.  The figure plots cumulative F770W mask boundaries for different maximum scales, starting at 100~pc (green), running through 140, 200, 280~pc (yellow, red, cyan), up to 400~pc (magenta). The top layer contour (100~pc, green) already \edit1{encapsulates the majority of the overall} structure of the DFN, but several regions of seemingly contiguous filamentary emission remain disconnected either internally or to the network. By looking at other colored contours emerging from under the green boundary, one can infer how adding progressively larger scales changes the network.  The 200~pc cumulative mask (red) appears to provide reliable recovery of all filamentary emission structures without undue peripheral excess, although 140~pc and 280~pc are probably also acceptable. \HST B-band filament extraction at 280 and 400~pc occasionally confuses interarm gaps as attenuation.  Such large ($>200$\,pc) scales also push the limit of what can be considered a filament in the sense of forming via turbulent Jeans scale gravitational instabilities (see \citealt{MEIDT_PHANGSJWST}).

Fig.~\ref{fig:dustmotes} presents a zoomed in view of a region east-southeast of the galaxy center, showing the filtered B-band data in comparison to an unprocessed version of the same image and to F1000W data from \JWST.  Contours of the 25~pc B-band and F1000W filaments are overplotted. This figure visually emphasizes the 40\% sight line overlap and morphological agreement at this scale.  Fig.~\ref{fig:dustmotes} also demonstrates feasibility to probe even smaller scales in visible dust lanes across the entire PHANGS-HST sample.  Significant substructuring (down to the $\sim5$~pc WFC3/UVIS resolution limit in NGC\,628) in attenuation of the features detected as dust emission filaments at 25~pc scales is apparent.
We further expect \HST to detect additional small scale filaments beyond the limit of MIRI -- several super narrow attenuation filaments without corresponding larger scale intensity depression are apparent in other regions of NGC\,628. Finally, Fig.~\ref{fig:dustmotes} demonstrates that with \HST we will be able to reveal candidate features we nickname `dust motes' (examples marked with yellow circles), essentially compact ($\lesssim10$~pc), dark clouds we cannot cleanly identify solely with \JWST due to confusion with point-like dusty extreme AGB stars \citep{Thilker_xAGB} lying outside the emission filament network (examples marked with green circles) in the short wavelength MIRI images. These dust motes could be individual molecular clouds in relative isolation.  Their size is comparable to the Taurus Molecular Cloud, although Taurus is star-forming whereas the motes may often be quiescent.  A better Milky Way \edit1{analog} might be the smallest scale clouds found in the 3D extinction maps of e.g. \citet{Leike2020}.  Dust mote clouds are challenging to confirm with MIRI, but appear in the \HST B-band images before any pre-processing (Fig.~\ref{fig:dustmotes} center). PHANGS-JWST F335M 3.3$\micron$ PAH imaging  \citep{SANDSTROM2_PHANGSJWST, RODRIGUEZ_PHANGSJWST} and forthcoming PHANGS-HST H$\alpha$ imaging (GO 17126, PI R.\ Chandar) may prove useful to further vet dust \edit1{mote} candidates as a class. 

\subsection{Fraction of flux in MIR filament network}
\label{sec:fluxfrac}
The relative amount of structured and unstructured (diffuse) dust in a galaxy is fundamental metric of ISM morphology and offers a basis for comparison with simulations that aim to understand the impact of factors such as stellar feedback and dynamical influences \citep[e.g.][]{Smith2020}. As noted in Sec.~\ref{sec:linktoSF}, this observable is also relevant to star formation rate (SFR) estimation. We measure the fraction of flux contained within the extracted filaments compared to the total flux in the image.  As the filaments are brighter than their surroundings, this requires that we establish a local background estimate at all locations in the image, which we do using the procedure given in Sec.~\ref{sec:analysis}. Figure~\ref{fig:fluxfrac} presents our results concerning flux fraction as a function of filament extraction scale. 

We find that at the smallest independent extraction scales considered (25~pc for F770W, F1000W, F1130W; 35~pc for F2100W), the background-subtracted filament flux fractions are nearly $30\%$ except for F2100W at $\approx20\%$.  The flux fraction is essentially constant versus extraction scale for F770W, F1000, and F1130W, but peaks at 35~pc and 200~pc for F2100W.  The F2100W flux fractions are always significantly less than all three shorter wavelength MIRI bands.  Given the present residual $\pm$\,0.1 MJy\,sr$^{-1}$ uncertainty in the background level of our MIRI images (see Sec.~\ref{sec:jwstdata} and \citealt{LEROY1_PHANGSJWST}), we assess the impact this has on filament flux fractions, by offsetting the image intensities by the uncertainty in a positive and negative sense.  Dashes overplotted on the bars of Fig.~\ref{fig:fluxfrac} indicate the resulting perturbed fractions. 

Cumulative multi-scale mask filament flux fraction results are also shown in Fig.~\ref{fig:fluxfrac}, indicated by the solid lines and dots.  As expected, the fraction increases as progressively more individual masks are combined and the maximum allowed extraction scale is increased.  For our choice of maximum scale (200~pc) the background-subtracted filament flux is in the range $55$\% to $60$\% of the total F770W, F1000W, and F1130W flux in the field.  At the same cumulative scale, the F2100W measurement is 43\%.  Scatter due to uncertainty in background level is also checked for cumulative values, and shown with dotted lines on Fig.~\ref{fig:fluxfrac}. The impact is notably larger than for individual scales, but is generally less than systematic uncertainty due to our choice of maximum scale included in the DFN cumulative mask.   

We note that if diffuse emission is neglected, rather than subtracted away as in our analysis, the integrated flux from the filament network becomes 40-100\% higher boosting the cumulative filament flux fractions to $\sim$80\% for F770W, F1000W, and F1130W and 60\% for F2100W. 

\subsection{Linking the dust filament network to star formation activity}
\label{sec:linktoSF}
The results of Sec.~\ref{sec:fluxfrac} confirm that more than half of the MIR flux from NGC\,628 originates in the intricately structured filament network, with the only exception being for F2100W (43\%). This outcome is what we anticipated based on previous work for nearby galaxies \citep{Liu2011, Leroy+2012, Crocker2013, Calzetti2013, Boquien2016, Kumari2020, Belfiore2022}.  It reinforces the necessity of correcting the integrated MIR luminosity of a galaxy for the presence of inherently diffuse emission heated by older stellar populations, when using MIR to measure current SFR \citep{LonsdalePersson1987, Boquien2016}. One of the goals of PHANGS-JWST is to clarify systematics of this correction. We can positionally test the linkage of the DFN to age-dated markers of star formation events, checking whether all regions of the filament network can be attributed to heating by current star formation or if some filamentary structure is effectively quiescent and should be removed alongside diffuse emission when estimating SFR.  Put differently, perhaps only filamentary structure up to a certain maximum scale, or down to a limiting dust surface density, is directly linked to the youngest stellar populations.

We take a first step toward the goal above by cross-correlating the filament masks with the PHANGS-MUSE \HII regions, plus PHANGS-HST young ($\le5$~Myr) clusters and associations, described in Sec.~\ref{sec:data}.  With these populations, we measure unweighted inclusion fractions (number of objects contained within the filament mask divided by the total count in the image footprint) and also tracer-weighted inclusion fractions (summing weights corresponding to SFR rather than object counts).  For weighting we use extinction corrected L(H$\alpha$), effectively unobscured SFR, for \HII region populations; and M$_*$ / age $\approx$ effective SFR(5 Myr) for young clusters and associations.

Fig.~\ref{fig:piegraph_InclFrac_vs_scale} plots the measured inclusion fractions for each of the test populations.  \edit1{As a control, we also evaluated band- and scale-dependent inclusion fractions (and their standard deviations) obtained for many sets of randomized positions in the footprint. Fig.~\ref{fig:piegraph_InclFrac_vs_scale} shows the randomized fractions in the top two panels.}
We begin by making general comments about Fig.~\ref{fig:piegraph_InclFrac_vs_scale} which are applicable to all panels.  We find that the inclusion fractions (at individual scales or cumulatively) are consistently maximized for \HII regions but decline  slightly when using young clusters and dramatically using young associations, regardless of scale or band.  The inclusion fraction measured at individual filament extraction scales for \HII regions and young clusters ($\le5$~Myr), decline from a maximum at 25~pc until reaching 70--100~pc.  For associations, there is far weaker, if any, dependence on scale over the same range, but an apparent enhancement at 140--200~pc scales.  The random inclusion fractions for unweighted measurements (top panels) provide further insight. They indicate strong correlation between dust emission filaments and \HII regions / young clusters, moderate anti-correlation of associations with filaments at scales $<$100~pc, and increasingly strong correlation of associations with the DFN for extraction scales of 100, 140, and 200~pc.  For measurements made with cumulative masks, the inclusion fractions (unweighted and weighted) increase with scale as would be expected.  The slope of the cumulative curves in Fig.~\ref{fig:piegraph_InclFrac_vs_scale} (top) is consistent with the scale dependence of the randomized control, showing the influence of significant multi-scale mask covering fraction, but normalization is significantly higher (excess over random $20\pm5\,\sigma$ for \HII regions, 3--6$\sigma$ for young clusters) except for associations.  For associations, the small scale signal of anti-correlation dominates until our cumulative measurement finally reaches random equivalence by 200~pc. 

The SFR-weighted inclusion fractions Fig.~\ref{fig:piegraph_InclFrac_vs_scale} (bottom) are perhaps more physically relevant to the questions raised at the start of this Section. Remarkably, our measurements show about 75\% of the SF traced by \HII regions is occurring within the 25 pc scale of the F770W filament network (bottom left), whereas the peak SFR-weighted inclusion fraction for F2100W is $\sim$45\% (at 35 pc, bottom right). We find the single scale SFR-weighted inclusion fractions are highest overall for F1130W and F1000W at nearly 80\% for 25~pc filament extraction (not shown). Perhaps the most striking measurement linking the dust emission filaments to current star formation is the attainment of $\sim95$\% SFR-weighted inclusion by our uppermost cumulative scale, 200~pc, for each of F770W, F1000W and F1130W when using the multi-scale masks, and 72\% for F2100W.  \edit1{Our finding of decreased inclusion fractions for F2100W agrees with assessment of relative suitability of MIRI bands as a gas column density tracer by \cite{LEROY1_PHANGSJWST} and \cite{SANDSTROM2_PHANGSJWST}, which show that the PAH-dominated bands are preferable to F2100W as a high-resolution proxy for gas.}

We anticipated seeing a reduction in inclusion fraction for stellar associations, since, although the associations are young, they are star formation events that have evolved from clusters via disruption/dissolution or were initially formed unbound, and either way seem more likely to have already cleared their environment of dust.  However, the anti-correlation observed at small scales is an unexpected demonstration that such feedback actively helps to sculpt the dust into bubble/shell structures \citep[e.g.][]{WATKINS_PHANGSJWST,BARNES_PHANGSJWST}.

Work is currently underway to obtain a catalog of PHANGS-ALMA GMCs associated with embedded ($t$\,$\sim$\,0 Myr) star formation, indicated by compact F2100W sources coincident with a GMC \citep{LessingThilker2023}.  When ready, we will compare our dust emission filaments to that population.  We emphasize that the F2100W source population does not drive the identification of filaments at $21\micron$ due to \textsc{FilFinder}'s transformation of the image intensity scale.

\section{SUMMARY AND FUTURE WORK}
\label{sec:summary}

This paper presents initial exploratory analysis of PHANGS-JWST+HST imaging for NGC\,628 revealing its extensive dust filament network (DFN) as seen in both MIR emission and visible attenuation.  Our pilot investigation offers insight into extragalactic ISM structure at small scales rarely probed by other tracers, including atomic and molecular gas, with an emphasis on quantifying filaments and associated star formation activity. 

Conclusions from our study are as follows:

\begin{enumerate}
    \item At the smallest extraction scale currently considered (25~pc filament width), the agreement of independently constructed attenuation and emission filament masks is 40\%.  More so, the detailed morphology of filaments that are detected in both ways is frequently rather well-matched with only minor deviations in shape or extent.  We find evidence for emission-only filaments (and portions of filaments) likely on the `back side' of the galaxy disk, but also a less well-understood set of attenuation-only filaments that requires further characterization.
    \item No single extraction scale (filament width) provides a complete inventory of all filamentary dust emission structures. Our \edit1{structure} masks are detecting a morphologically diverse dusty ISM, spanning from very compact \edit1{GMC-sized} filaments to \edit1{larger scale, lower surface brightness filaments} ultimately constituting \edit1{galactic} spiral arm features.  We anticipate that the molecular fraction probably declines with increasing scale in this hierarchy, but require higher resolution observations of atomic gas in order to check this.  
    \item \HST reveals candidate features we nickname `dust motes’ which are comparatively isolated (lying outside the DFN) and appear as compact ($\lesssim10$~pc), dark clouds essentially unrecoverable with \JWST/MIRI alone due to confusion with dusty stellar point sources.  They could trace largely quiescent individual molecular clouds.  Regardless of whether candidate `dust motes' are ultimately verified as a bona fide population, HST is capable of probing substructure in dust filaments at smaller scales than MIRI.
    \item Approximately one-third of the total MIR flux in F770W, F1000W, F1130W bands is contained in the 25~pc scale mask of the emission filament network, using diffuse background subtracted measurements.  The flux fraction determined for the 200~pc limited cumulative multi-scale filament mask is 55--60\% in the same bands.  The F2100W filament flux fractions are significantly less than the others, with a cumulative measurement of $\sim45$\%.  This is in-line with \cite{LEROY1_PHANGSJWST} who showed that F2100W correlates less well with CO than the other MIRI bands and that it is not as clean of a tracer of column density.
    \item Our filament inclusion fraction analysis shows that 75--80\% of the current star formation traced by \HII regions is occurring within the 25 pc scale of the filament network.  The analogous measurement for young ($<$5~Myr) clusters is slightly more than 60\%.  Integrated over cumulative scales up to 200 pc, the \HII region fraction exceeds 95\%.  Similar analysis demonstrates moderate anti-correlation of associations younger than 5 Myr with dust filaments at scales $<$100~pc then a reversal to increasingly strong association-filament correlation for extraction scales of 100, 140, and 200~pc.  
\end{enumerate}

Expansion and further development of our work to the remaining PHANGS-JWST galaxies will: (a) provide the clarity of an external and diversified perspective which is absent from Galactic studies, and which can allow quantification of trends \edit1{in filament properties with local environmental physical conditions}, (b) enable comparison to increasingly detailed simulations, which can isolate the effects of different ISM structuring mechanisms, (c) constrain the dependence of opacity on dust grain properties, and eventually (d) firmly quantify the division of MIR emission between currently star-forming and evolved stellar populations \citep[related to the work of][]{Belfiore2022}, with complete accounting of unobscured \textit{and embedded} star formation \citep{RODRIGUEZ_PHANGSJWST,HASSANI_PHANGSJWST}. The union of \JWST and \HST dust tracing also motivates targeting for focused high-resolution ALMA follow-up mapping of dense gas. We further emphasize that ngVLA and/or SKA is required to obtain sensitive \HI imaging at the substantially better than $6\arcsec$ resolution that is needed to constrain ISM phase changes within the filamentary structures we study.

We plan the following practical improvements to our work in the near term.  Attenuation (dust lane) features will be identified on the basis of the complete multi-wavelength PHANGS-HST dataset and  at sub-MIRI-resolution scales. \edit1{Determination of attenuation, A(V), as a function of position \citep[e.g.][]{Viaene2017, ButlerTan2009} within the filament network will be accomplished, allowing pixel-by-pixel correlation of emission and attenuation to be measured.} 
We will study the integrated SED of the filament network versus scale, and use \textsc{FilFinder} capabilities to generate catalogs of filament substructures with measured properties (such as length, aspect ratio, curvature, flux, mass per unit length) plus molecular-cloud-linked quantities (velocity gradient, CO velocity dispersion, virial parameter) when a GMC is found to be co-spatial. Such cataloged properties will be ripe for comparison to equivalent quantities measured from simulations. We will assess whether or not embedded star-forming regions tend to be located at the intersection of filament spines, as they do in simulations. \edit1{Lastly, we will analyse ensemble characteristics such as those noted above} of dust emission and attenuation filament network substructures versus physical condition metrics \citep[e.g.][]{sun2022} of the local ($\sim$kpc-scale) environment.

\section*{Acknowledgements}

This work was carried out as part of the PHANGS collaboration. 
Based on observations made with the NASA/ESA/CSA JWST and Hubble Space Telescopes. The data were obtained from the Mikulski Archive for Space Telescopes at the Space Telescope Science Institute, which is operated by the Association of Universities for Research in Astronomy, Inc., under NASA contract NAS 5-03127 for JWST and NASA contract NAS 5-26555 for HST. The JWST observations are associated with program 2107, and those from HST with program 15454
Based on observations collected at the European Southern Observatory under ESO programmes 094.C-0623 (PI: Kreckel), 095.C-0473,  098.C-0484 (PI: Blanc), 1100.B-0651 (PHANGS-MUSE; PI: Schinnerer), as well as 094.B-0321 (MAGNUM; PI: Marconi), 099.B-0242, 0100.B-0116, 098.B-0551 (MAD; PI: Carollo) and 097.B-0640 (TIMER; PI: Gadotti).
We acknowledge the usage of the SAO/NASA Astrophysics Data System\footnote{\url{http://www.adsabs.harvard.edu}}.
D.A.T. acknowledges funding support from STScI via JWST-GO-02107.002-A.
E.W.K. acknowledges support from the Smithsonian Institution as a Submillimeter Array (SMA) Fellow and the Natural Sciences and Engineering Research Council of Canada.
J.M.D.K. gratefully acknowledges funding from the European Research Council (ERC) under the European Union's Horizon 2020 research and innovation programme via the ERC Starting Grant MUSTANG (grant agreement number 714907). 
COOL Research DAO is a Decentralized Autonomous Organization supporting research in astrophysics aimed at uncovering our cosmic origins.
E.J.W. acknowledges the funding provided by the Deutsche Forschungsgemeinschaft (DFG, German Research Foundation) -- Project-ID 138713538 -- SFB 881 (``The Milky Way System'', subproject P1). 
M.C. gratefully acknowledges funding from the DFG through an Emmy Noether Research Group (grant number CH2137/1-1).
M.B. acknowledges support from FONDECYT regular grant 1211000 and by the ANID BASAL project FB210003.
T.G.W. and E.S. acknowledge funding from the European Research Council (ERC) under the European Union’s Horizon 2020 research and innovation programme (grant agreement No. 694343).
E.R. acknowledges the support of the Natural Sciences and Engineering Research Council of Canada (NSERC), funding reference number RGPIN-2022-03499.
F.B. would like to acknowledge funding from the European Research Council (ERC) under the European Union’s Horizon 2020 research and innovation programme (grant agreement No.726384/Empire)
K.G. is supported by the Australian Research Council through the Discovery Early Career Researcher Award (DECRA) Fellowship DE220100766 funded by the Australian Government. 
K.G. is supported by the Australian Research Council Centre of Excellence for All Sky Astrophysics in 3 Dimensions (ASTRO~3D), through project number CE170100013. 
R.S.K. acknowledges financial support from the European Research Council via the ERC Synergy Grant ``ECOGAL'' (project ID 855130), from the Deutsche Forschungsgemeinschaft (DFG) via the Collaborative Research Center ``The Milky Way System''  (SFB 881 -- funding ID 138713538 -- subprojects A1, B1, B2 and B8) and from the Heidelberg Cluster of Excellence (EXC 2181 - 390900948) ``STRUCTURES'', funded by the German Excellence Strategy. R.S.K. also thanks the German Ministry for Economic Affairs and Climate Action for funding in the project ``MAINN'' (funding ID 50OO2206). 
G.A.B. acknowledges the support from ANID Basal project FB210003.
M.Q. acknowledges support from the Spanish grant PID2019-106027GA-C44, funded by MCIN/AEI/10.13039/501100011033.
E.C. acknowledges support from ANID Basal projects ACE210002 and FB210003.
S.D. is supported by funding from the European Research Council (ERC) under the European Union’s Horizon 2020 research and innovation programme (grant agreement no. 101018897 CosmicExplorer).
K.K. and O.E. gratefully acknowledge funding from DFG in the form of an Emmy Noether Research Group (grant number KR4598/2-1, PI Kreckel).

\facilities{
\HST, \JWST, VLT:Yepun}

\software{\textsc{astropy} \citep{Astropy+2013,Astropy+2018},  \textsc{numpy} \citep{Harris+2020}, \textsc{matplotlib} \citep{Hunter+2007} and \textsc{FilFinder} \citep[v1.7.2;][]{KochRosolowsky2015}}

\restartappendixnumbering
\appendix

\section{Dust filament masks versus band and extraction scale}
\label{sec:appendix}
In to permit examination of the extracted dust filaments with respect to the \JWST and \HST data, we present the filament masks as a Figure Set. They are displayed first for F770W, then F2100W, and lastly HST B-band.  We show the filaments as transparent colored areas on the associated \JWST or \HST image.  For each band, we present all of the individual scale filament masks in order of increasing extraction scale, followed by the cumulative multi-scale masks.

\figsetstart
\figsetnum{1}
\figsettitle{Dust filament masks}

\figsetgrpstart
\figsetgrpnum{1.1}
\figsetgrptitle{F770W, 25~pc}
\figsetplot{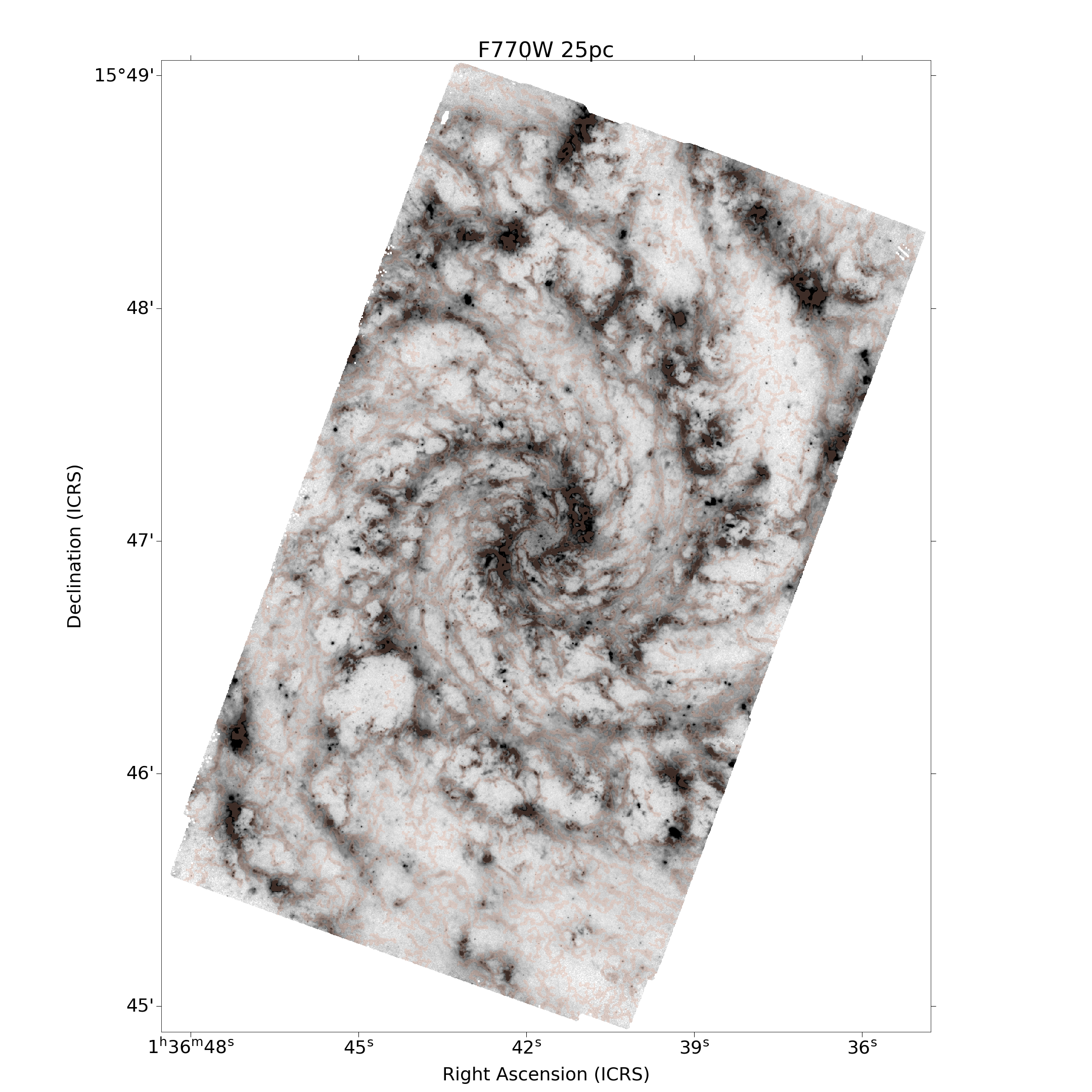}
\figsetgrpnote{Dust emission filament mask for F770W, 25~pc extraction scale.  The image is oriented with North up and East left.}
\figsetgrpend

\figsetgrpstart
\figsetgrpnum{1.15}
\figsetgrptitle{F770W, cumulative $<$200~pc}
\figsetplot{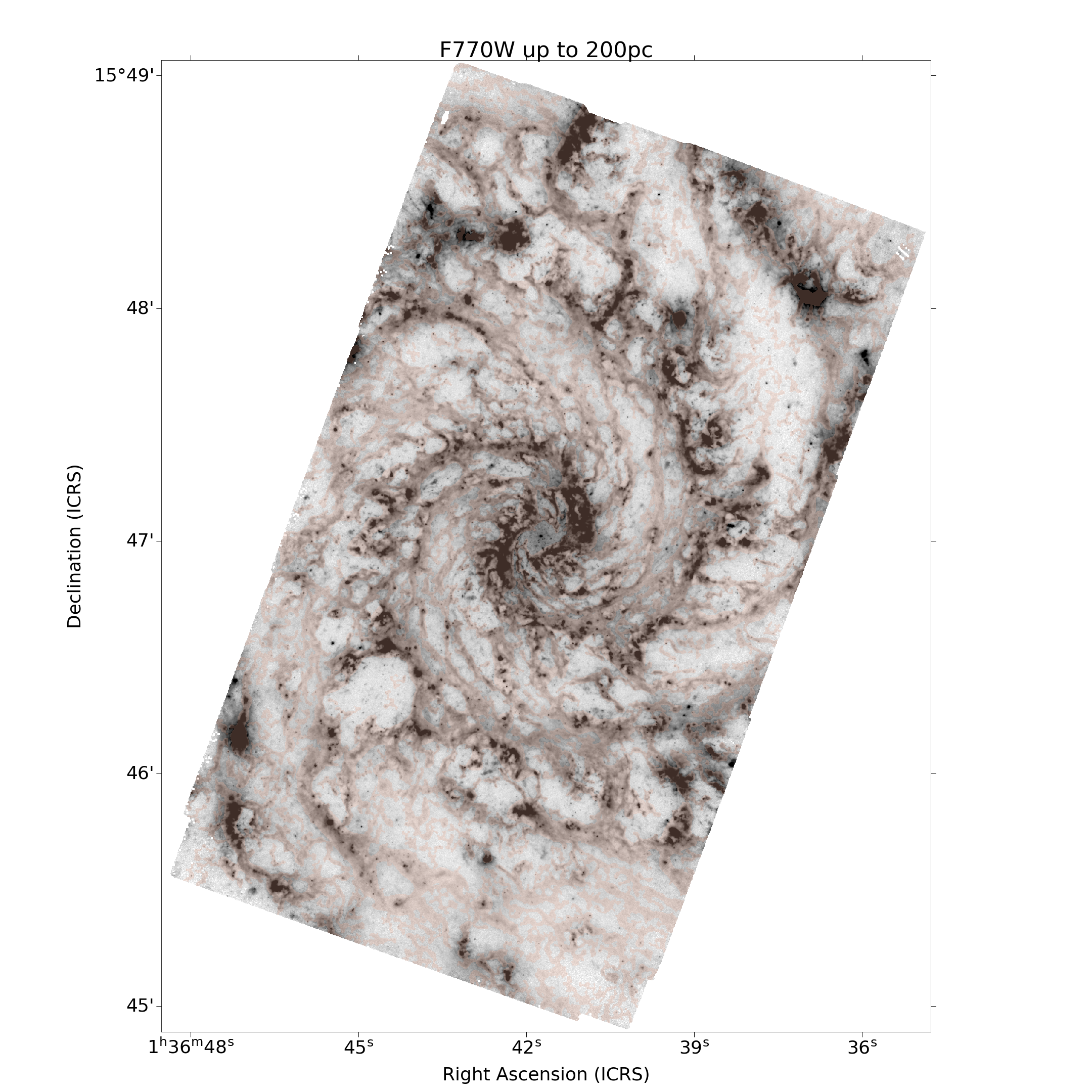}
\figsetgrpnote{Dust emission filament mask for F770W, cumulative $<$200~pc extraction scale.  The image is oriented with North up and East left.}
\figsetgrpend

\figsetgrpstart
\figsetgrpnum{1.16}
\figsetgrptitle{F2100W, 35~pc}
\figsetplot{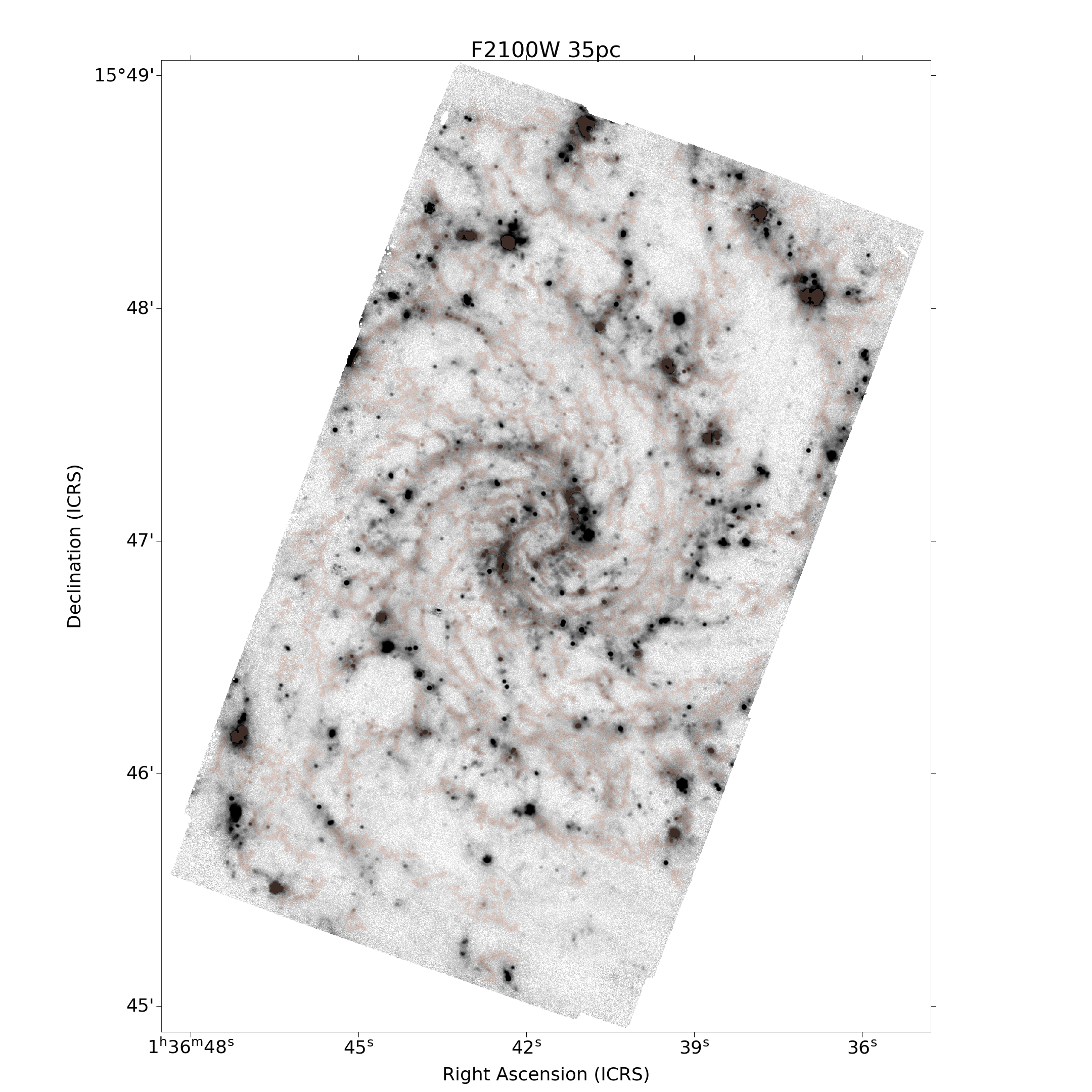}
\figsetgrpnote{Dust emission filament mask for F2100W, 35~pc extraction scale.  The image is oriented with North up and East left.}
\figsetgrpend

\figsetgrpstart
\figsetgrpnum{1.29}
\figsetgrptitle{F2100W, cumulative $<$200~pc}
\figsetplot{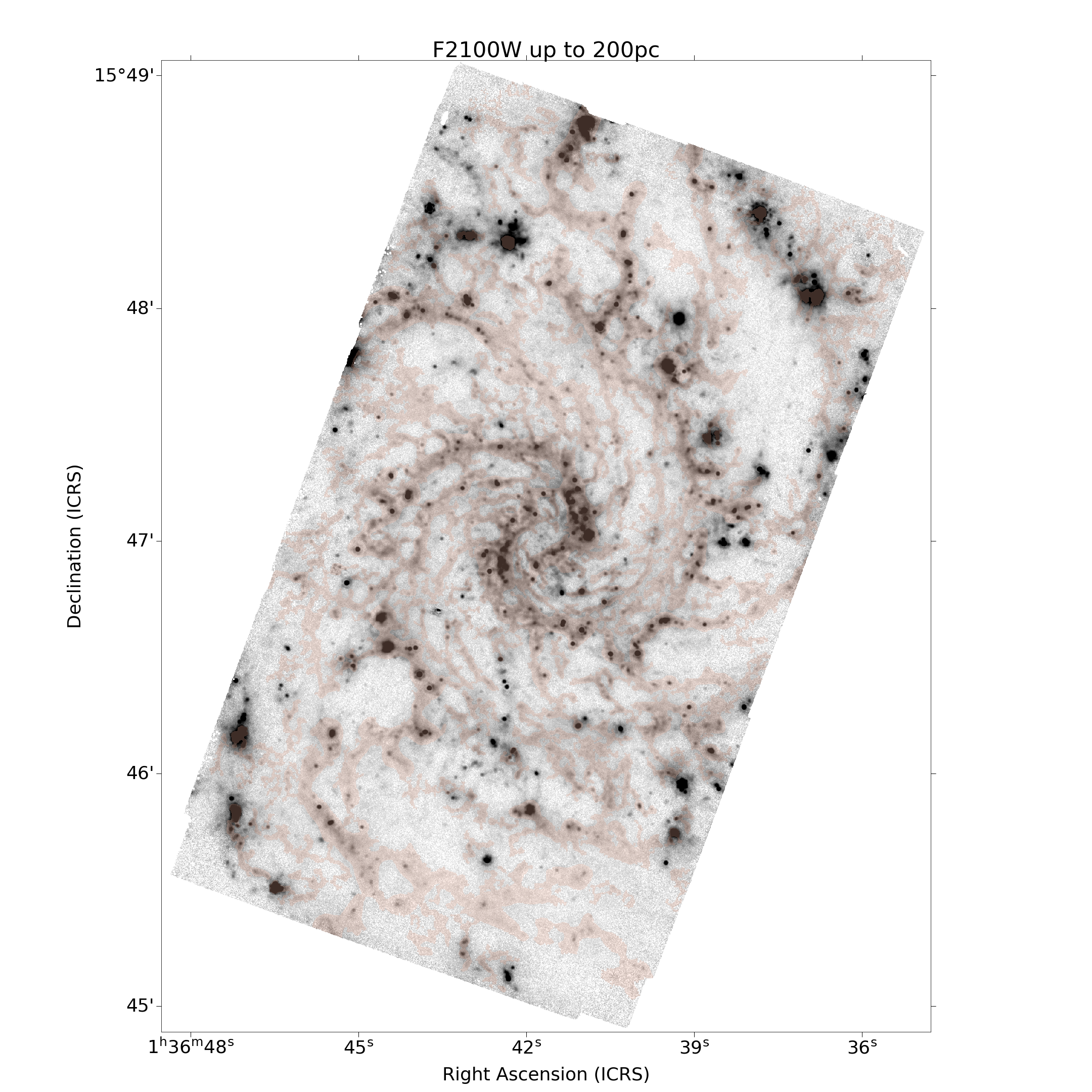}
\figsetgrpnote{Dust emission filament mask for F2100W, cumulative $<$200~pc extraction scale.  The image is oriented with North up and East left.}
\figsetgrpend

\figsetgrpstart
\figsetgrpnum{1.30}
\figsetgrptitle{HST B-band, 25~pc}
\figsetplot{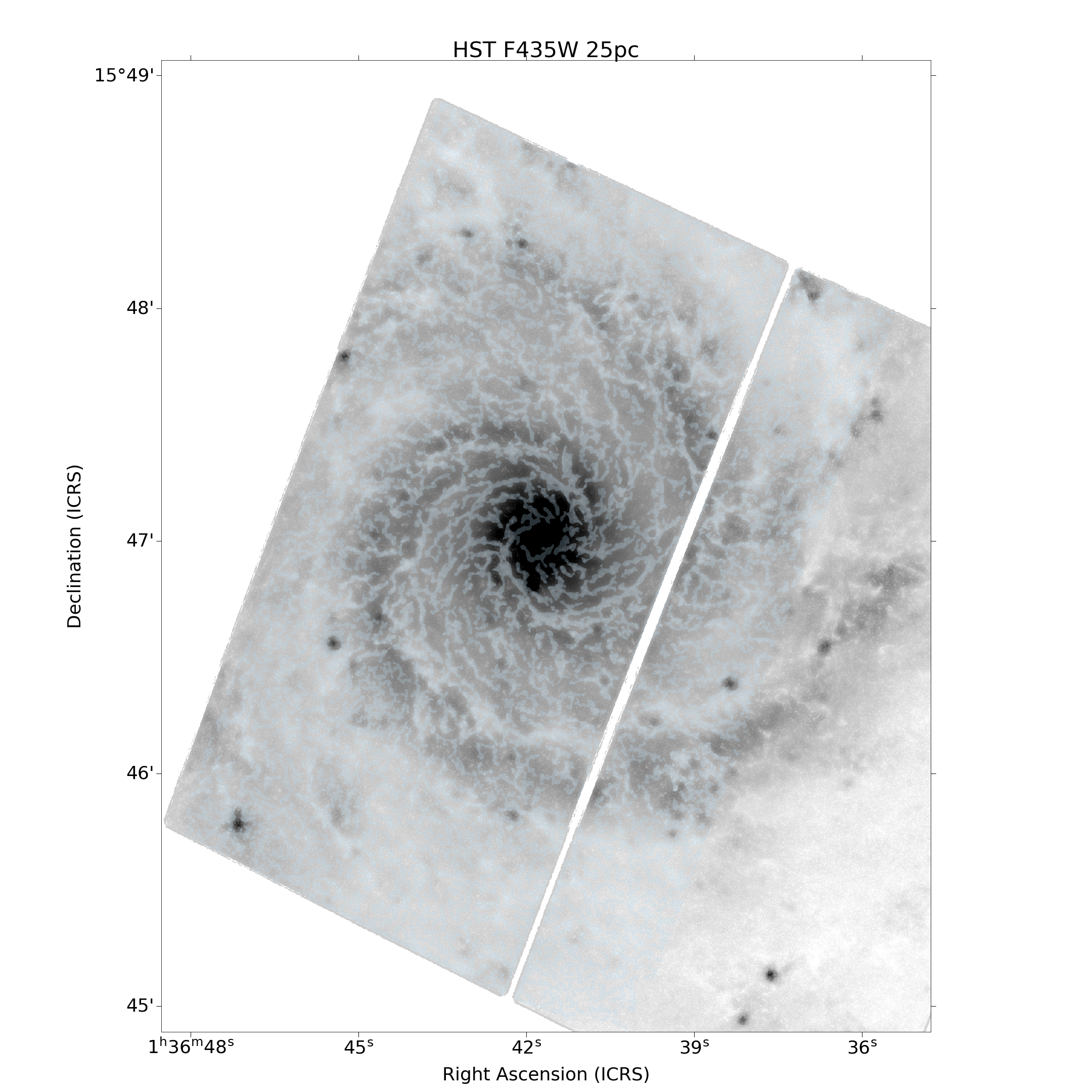}
\figsetgrpnote{Dust attenuation filament mask for \HST B-band, 25~pc extraction scale.  The image is oriented with North up and East left.}
\figsetgrpend

\figsetgrpstart
\figsetgrpnum{1.44}
\figsetgrptitle{HST B-band, cumulative $<$200~pc}
\figsetplot{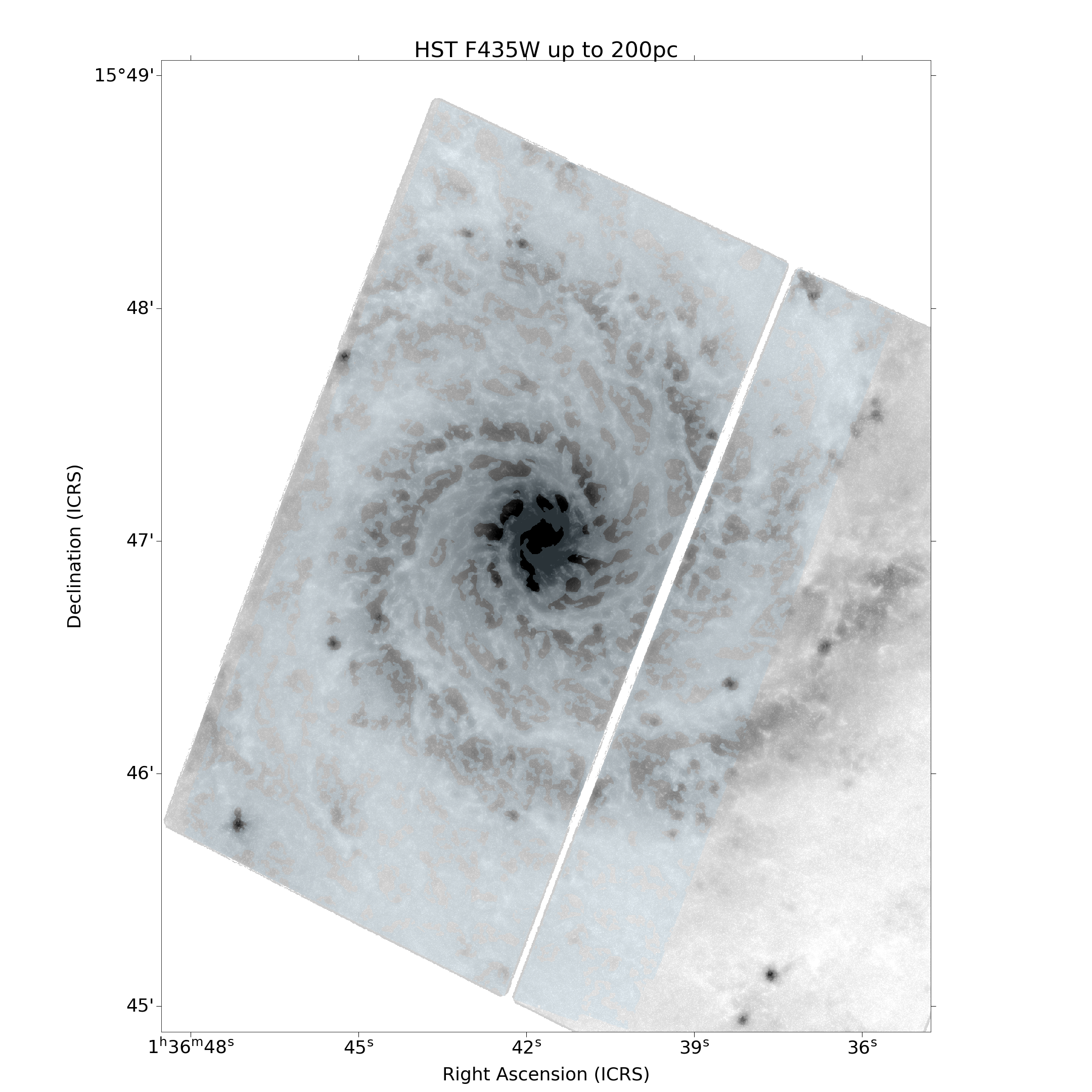}
\figsetgrpnote{Dust attenuation filament mask for \HST B-band, cumulative $<$200~pc extraction scale.  The image is oriented with North up and East left.}
\figsetgrpend

\figsetend

\begin{figure*}
\includegraphics[width=\linewidth]{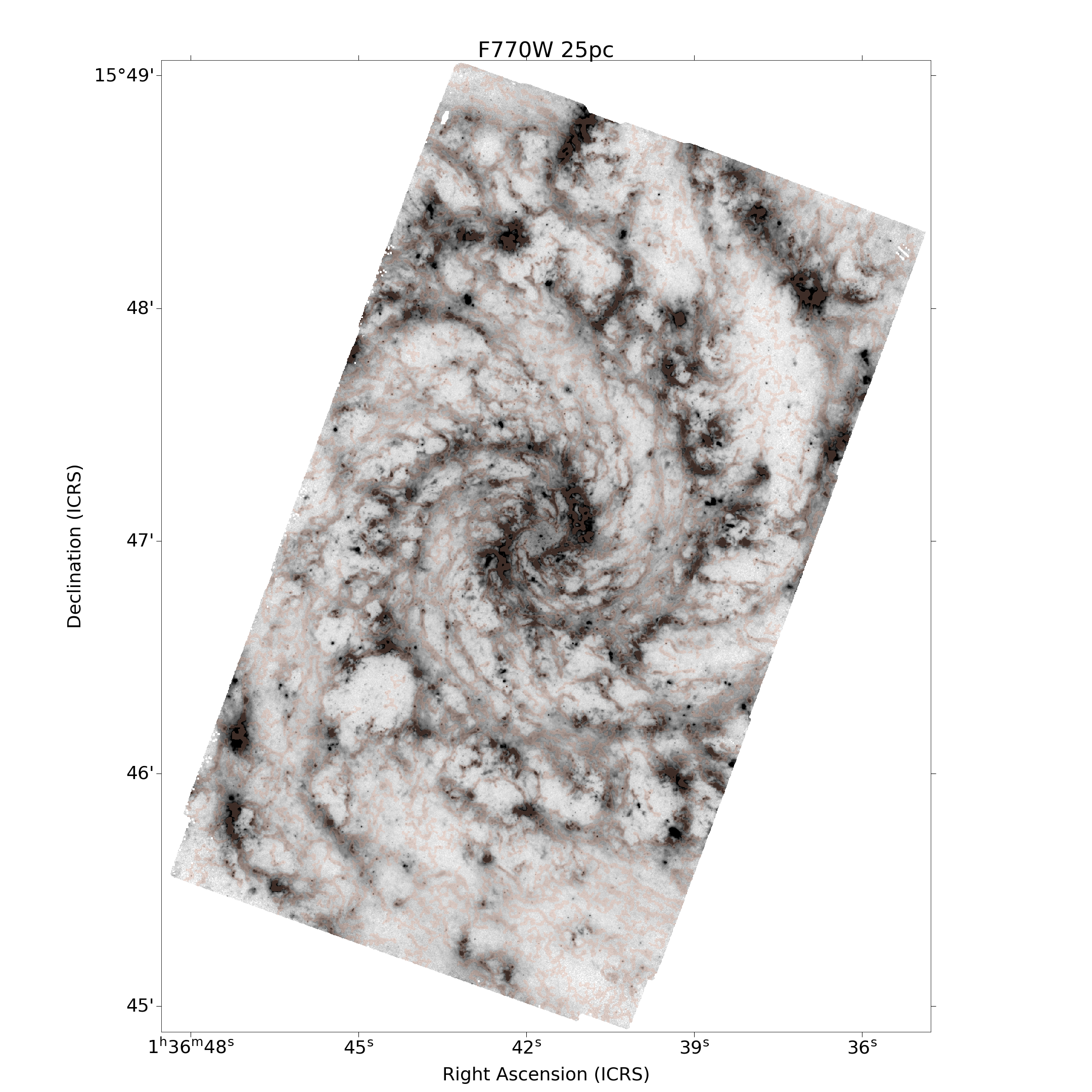}
\caption{\label{fig:appendixmasks1} \JWST F770W image with extracted emission filaments (25~pc scale) shown as transparent colored areas.  The image is oriented with North up and East left. The complete figure set (44 images) is available in the online journal\edit1{.}}
\end{figure*}  

\begin{figure*}
\includegraphics[width=\linewidth]{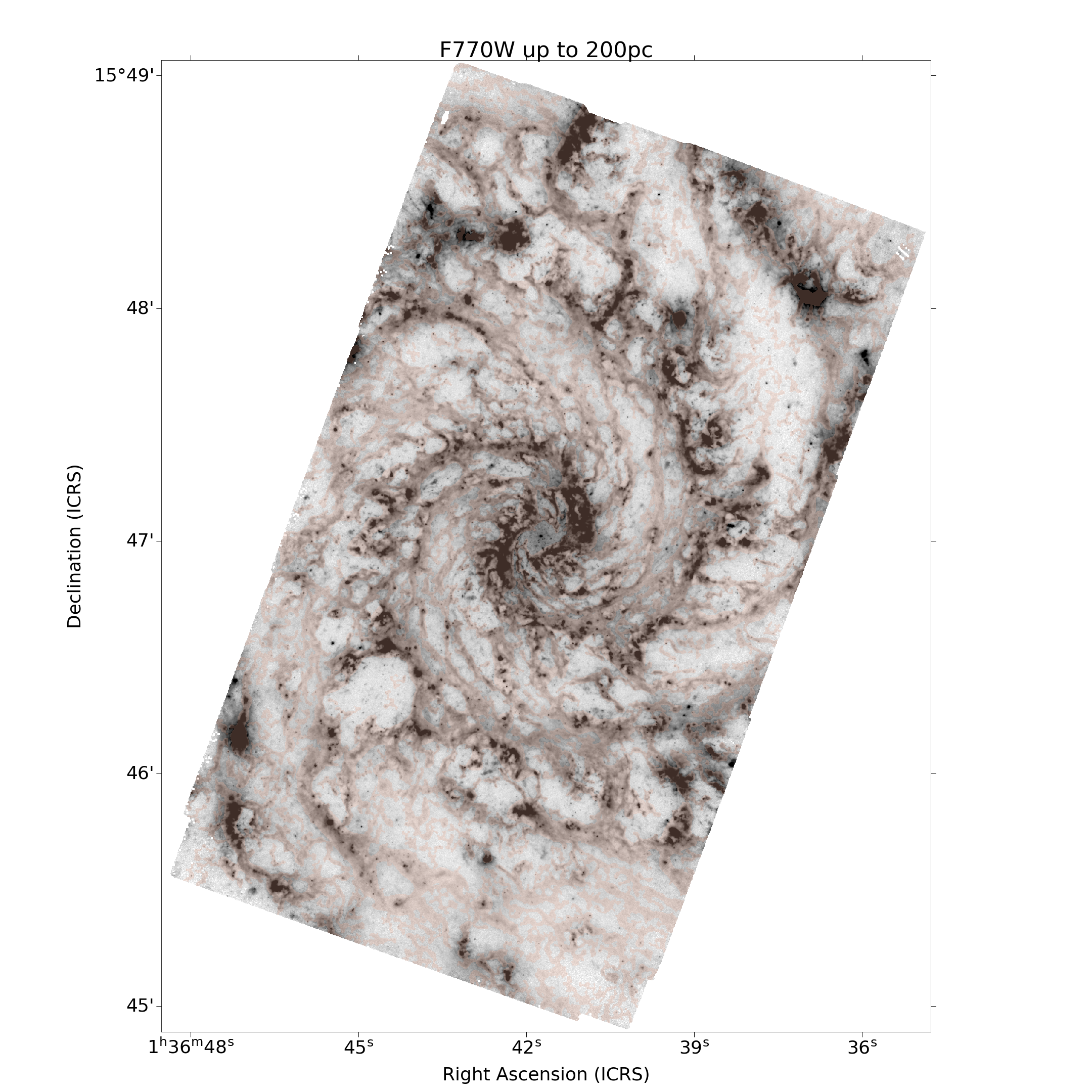}
\caption{\label{fig:appendixmasks2} \JWST F770W image with extracted emission filaments (cumulative up to 200~pc scale) shown as transparent colored areas.  The image is oriented with North up and East left. The complete figure set (44 images) is available in the online journal\edit1{.}}
\end{figure*}  

\begin{figure*}
\includegraphics[width=\linewidth]{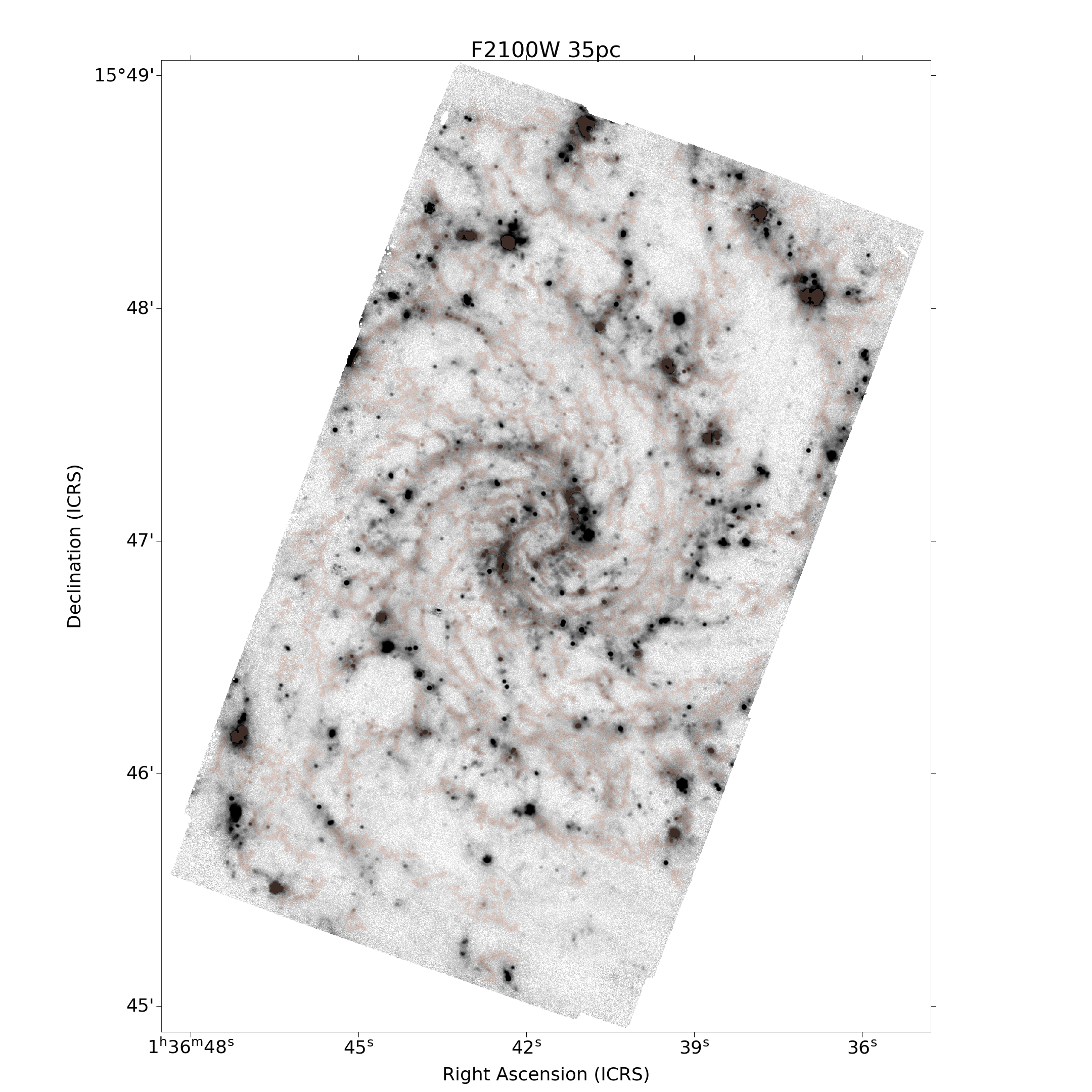}
\caption{\label{fig:appendixmasks3} \JWST F2100W image with extracted emission filaments (35~pc scale) shown as transparent colored areas.  The image is oriented with North up and East left. The complete figure set (44 images) is available in the online journal\edit1{.}}
\end{figure*}  

\begin{figure*}
\includegraphics[width=\linewidth]{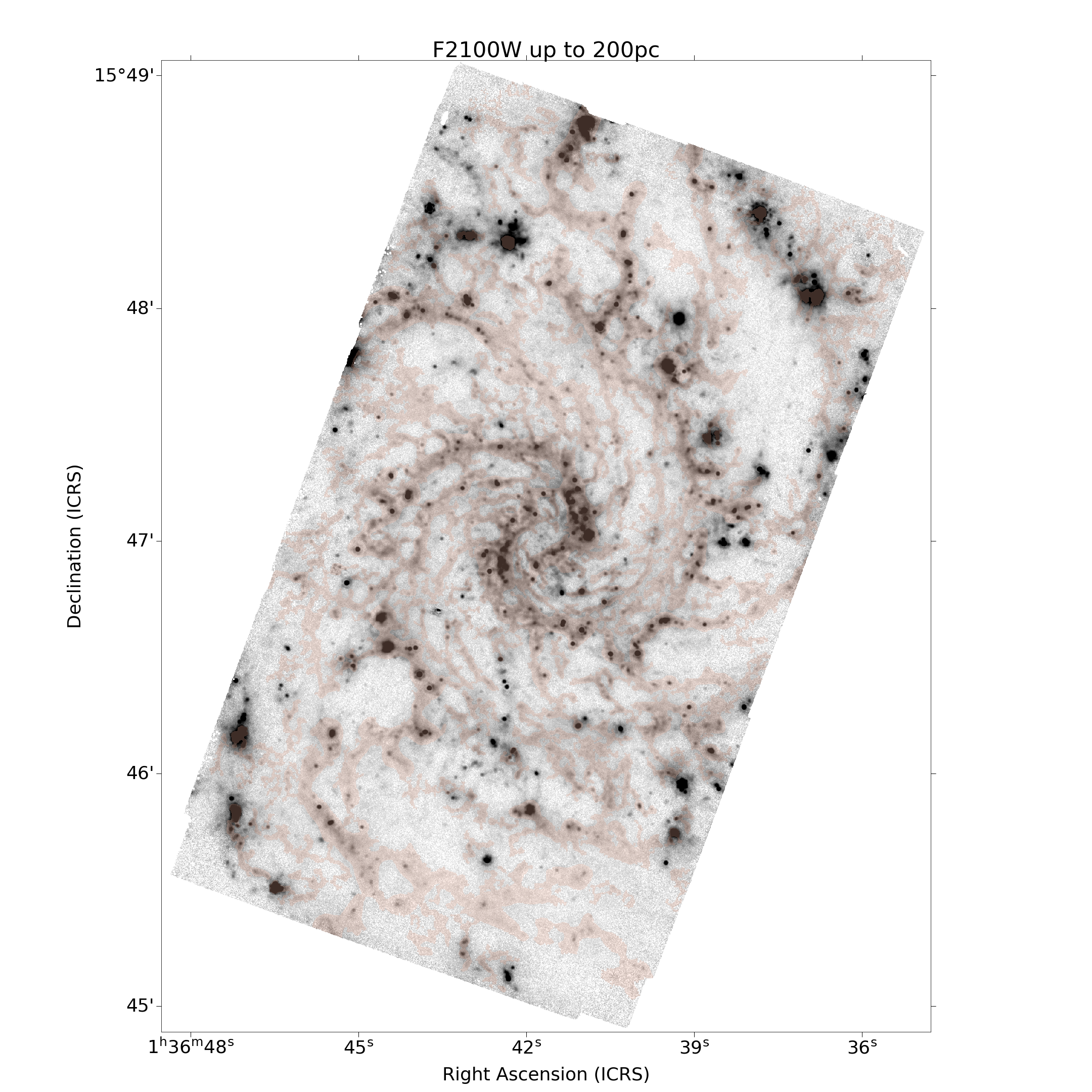}
\caption{\label{fig:appendixmasks4} \JWST F2100W image with extracted emission filaments (cumulative up to 200~pc scale) shown as transparent colored areas.  The image is oriented with North up and East left. The complete figure set (44 images) is available in the online journal\edit1{.}}
\end{figure*}  

\begin{figure*}
\includegraphics[width=\linewidth]{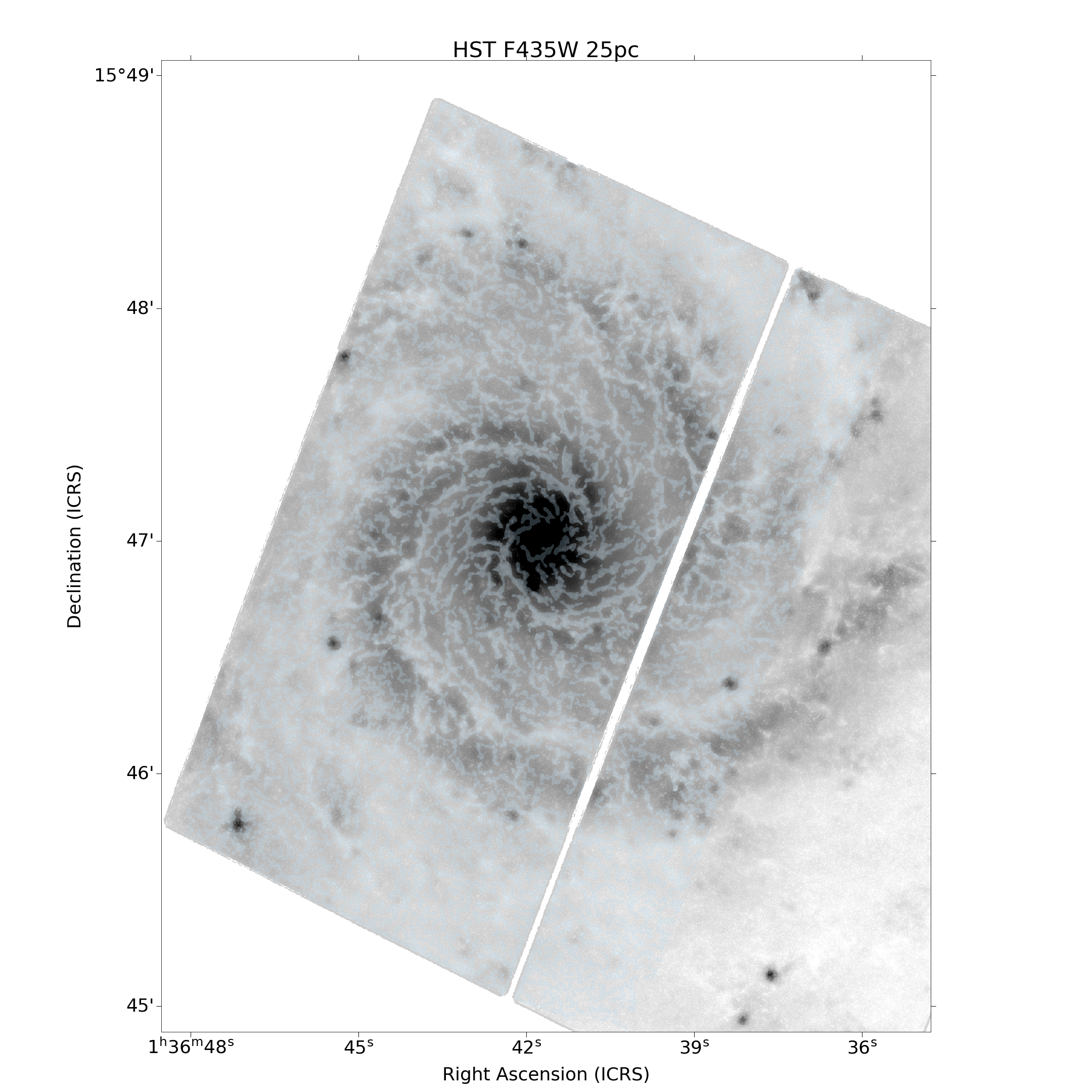}
\caption{\label{fig:appendixmasks5} \HST B-band (F435W) image with extracted attenuation filaments (25~pc scale) shown as transparent colored areas.  The image is oriented with North up and East left. The complete figure set (44 images) is available in the online journal\edit1{.}}
\end{figure*}  

\begin{figure*}
\includegraphics[width=\linewidth]{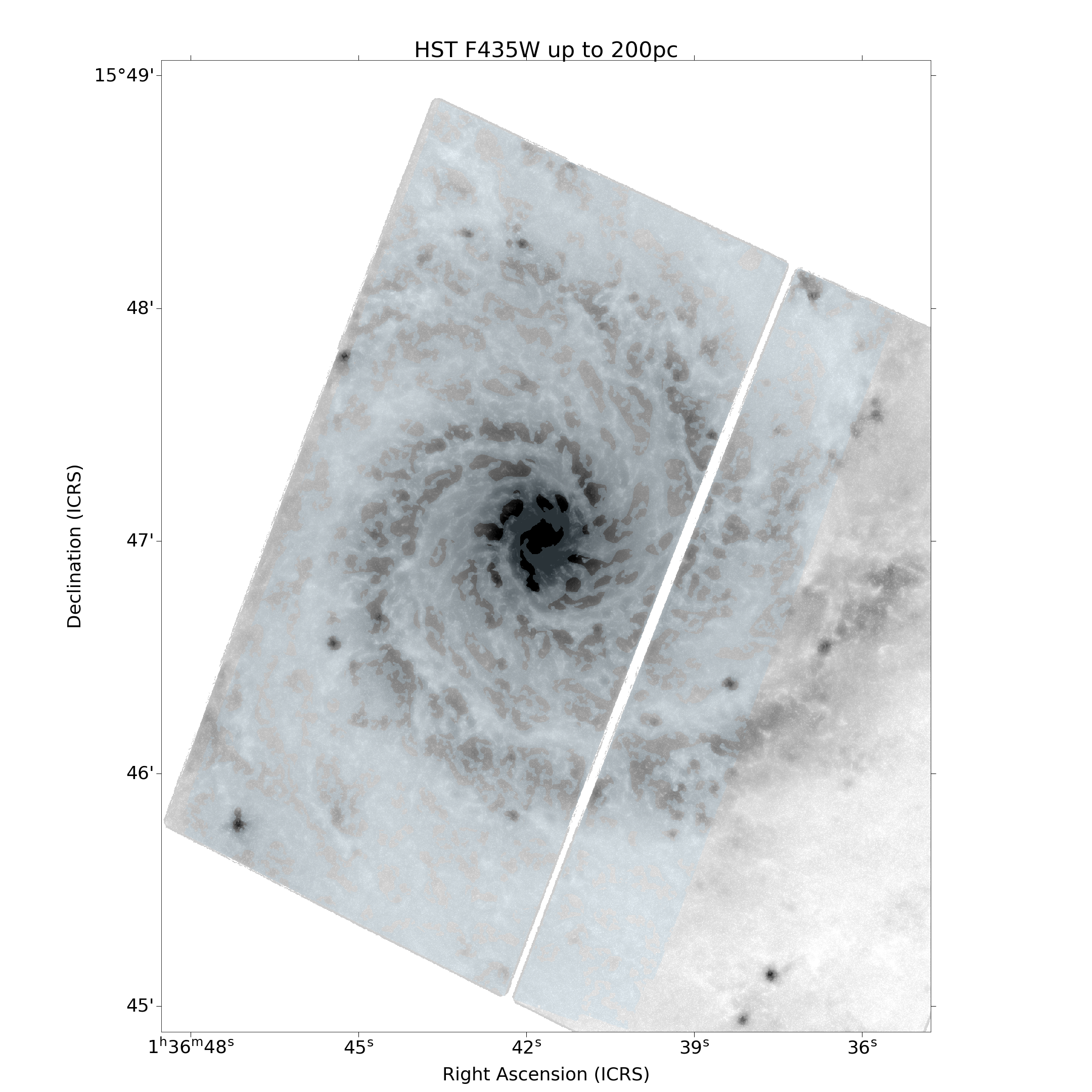}
\caption{\label{fig:appendixmasks6} \HST B-band (F435W) image with extracted attenuation filaments (cumulative up to 200~pc scale) shown as transparent colored areas.  The image is oriented with North up and East left. The complete figure set (44 images) is available in the online journal\edit1{.}}
\end{figure*}

\bibliographystyle{aasjournal}
\bibliography{paper,phangsjwst} 

\suppressAffiliationsfalse
\allauthors

\end{document}